\documentclass[aps,prc,preprint,superscriptaddress,nofootinbib]{revtex4-2}

\usepackage{amsmath,amssymb}
\usepackage{graphicx}
\usepackage{xcolor}
\usepackage[colorlinks=true,linkcolor=blue!50!black,citecolor=blue!50!black,urlcolor=blue!50!black]{hyperref}

\newcommand{\dd}{\mathrm{d}}
\newcommand{\lH}{\lambda_{\rm H}}
\newcommand{\Teq}{n_{\rm eq}}
\newcommand{\lPF}{\texorpdfstring{\mbox{$\lambda$-PF}}{lambda-PF}}

\makeatletter
\def\paragraph{\@startsection{paragraph}{4}{\z@}{3.25ex \@plus1ex \@minus.2ex}{-1em}{\normalfont\normalsize\bfseries}}
\makeatother
\begin{document}

\title{Accelerating exact solutions of relativistic hydrodynamics\\ and L\'evy-shaped freeze-out sources}

\author{Máté Csanád}
\author{Márton Nagy}
\author{Bence Szentkuti}
\author{Ágoston Benyó}
\affiliation{ELTE Eötvös Loránd University, Institute of Physics and Astronomy, H-1117 Budapest, Pázmány Péter sétány 1/A}

\date{\today}

\begin{abstract}
Femtoscopic measurements in high-energy heavy-ion collisions indicate that the pion-emitting source is L\'evy-stable rather than Gaussian, while the source radii ($R$) still follow the hydrodynamic scaling with transverse mass, $R^2\propto 1/m_T$. We call this coexistence the L\'evy hydro puzzle. For a perfect fluid with $\varepsilon=\kappa p$ and $p=nT$, flow and pressure form a closed system, and density and temperature can be ``dressed'' by an arbitrary comoving function; this generalises a freedom known from Hubble-type solutions to arbitrary flows. We show that, at nonzero pair transverse momentum, the source is bounded by the pressure on the freeze-out hypersurface, so the source tail can never be heavier than the pressure tail. This yields a general criterion for power-law source tails, with two simple routes. Either the flow in the outer regions becomes ultra-relativistic and the temperature grows at least linearly (Hubble-type solutions, where the fugacity in the Cooper--Frye formula suppresses the emission from the hot tail), or the flow is slow, the temperature uniform, and the proper acceleration falls off with distance as $1/r$. We present a new family of relativistic, accelerating, self-similar potential flows, which we call $\lambda$ potential flows ($\lambda$-PF), and a new dressing of Gubser's flow. The $\lambda$-PF family is regular, and its acceleration does not generate power-law tails; with a Hubble-type dressing, however, its source tail is L\'evy-stable in a range of parameters, which generalises the Hubble-type L\'evy source to accelerating cores. We show that in this route a L\'evy-stable tail requires an infinite particle number. The isothermally dressed Gubser flow realises the $1/r$-accelerating route. Its source tail is a power law with an exponent outside the L\'evy-stable range, fixed by the Gubser solution itself, but over the range probed by measurements it gives sources that are L\'evy-like in the transverse directions, with a L\'evy exponent $\alpha\approx1.7$ and $1/R^2$ linear in $m_T$, also for a weaker transverse flow that gives nearly realistic spectra; in the longitudinal direction the source is nearly Gaussian. The L\'evy-like shape is produced by the dilute outer region of the source. Our answer to the puzzle is that L\'evy-like sources and the hydrodynamic $m_T$ scaling are compatible in perfect-fluid hydrodynamics, but hydrodynamics does not create the power-law tail: the tail is either inherited from a power-law pressure, which requires a $1/r$ acceleration, or put in through the dressing. Neither new solution is fully three-dimensional, and we outline extensions towards non-spherical geometries and elliptic flow.
\end{abstract}

\maketitle

%=====================================================================
\section{Introduction: the L\'evy hydro puzzle}
\label{sec:intro}
%=====================================================================

The matter created in high-energy heavy-ion collisions behaves as a nearly perfect fluid, and relativistic hydrodynamics describes its expansion remarkably well~\cite{Heinz:2013th,Ollitrault:2007du,Florkowski:2010zz}. The space--time structure of the fluid at kinetic freeze-out is probed by femtoscopy, i.e.\ Bose--Einstein (HBT) correlations of identical particles~\cite{Lisa:2005dd}. Two observations from these measurements are in tension.

\begin{enumerate}
\item The source size parameters follow the scaling $R^{-2}\simeq a+b\,m_T$, which is characteristic of hydrodynamic expansion~\cite{Makhlin:1987gm,Csorgo:1995bi}.
\item The source shape is not Gaussian but is well described by a L\'evy-stable distribution with index $\alpha<2$~\cite{Csorgo:2003uv,PHENIX:2017ino,NA61SHINE:2023qzr,CMS:2023xyd}.
\end{enumerate}

L\'evy-stable distributions are the natural outcome of anomalous diffusion (L\'evy flights)~\cite{Metzler:2000qhw,Csanad:2007fr,Kincses:2024lnv}. If the non-Gaussian shape were produced by a non-hydrodynamic random process, it would be surprising for the hydrodynamic $m_T$ scaling to survive. Conversely, standard hydrodynamic sources with a thermal (Boltzmann) cutoff tend to be Gaussian-like. We refer to the question of how a hydrodynamically evolving fluid can produce a L\'evy-shaped source while keeping the hydrodynamic scaling of its radii as the \emph{L\'evy hydro puzzle}.

Exact solutions of hydrodynamics are a natural tool to address this question. Every step can be followed analytically, the role of each parameter is transparent, and the solutions serve as benchmarks for numerical codes. Our aim is to identify exact solutions that are (i) relativistic, (ii) accelerating, (iii) genuinely 1+3 dimensional, and (iv) produce realistic observables including L\'evy-like sources. To our knowledge no solution satisfying all four requirements has been found so far. The solutions presented here do not satisfy all four either; Table~\ref{tab:requirements} in Sec.~\ref{sec:summary} lists which requirements each of them meets.

The paper is organised as follows. Section~\ref{sec:exp} summarises the experimental status of L\'evy femtoscopy, and Sec.~\ref{sec:framework} the hydrodynamic and freeze-out framework. Section~\ref{sec:hubble} reviews Hubble-type solutions, their successes and their problems, and how the fugacity in the freeze-out resolves the latter. Section~\ref{sec:history} reviews accelerating solutions. Section~\ref{sec:general} contains general results: the dressing theorem, the bound of the source by the pressure, and the tail criteria. Section~\ref{sec:solI} presents a new family of accelerating potential flows, and Sec.~\ref{sec:solII} a new dressing of Gubser's flow. Section~\ref{sec:obs} computes observables, Sec.~\ref{sec:nonsph} discusses non-spherical extensions, and Sec.~\ref{sec:summary} summarises and states our answer to the puzzle.

%=====================================================================
\section{Experimental status of L\'evy femtoscopy}
\label{sec:exp}
%=====================================================================

\subsection{L\'evy-stable sources}
For identical bosons with average momentum $K$ and relative momentum $Q$, the two-particle correlation function is in the absence of final-state interactions, to a good approximation,
\begin{equation}
C(Q,K)\simeq1+\lambda(K)\,\frac{|\tilde S(Q,K)|^2}{|\tilde S(0,K)|^2},\label{eq:CQ}
\end{equation}
where $\tilde S$ is the Fourier transform of the source function and $\lambda$ the intercept (core--halo) parameter~\cite{Lisa:2005dd}. A spherically symmetric L\'evy-stable source of scale $R$ and index $\alpha$ is defined by its characteristic function~\cite{Csorgo:2003uv},
\begin{equation}
S(\mathbf r)=\int\frac{\dd^3q}{(2\pi)^3}\,e^{i\mathbf q\cdot\mathbf r}\,e^{-\frac12|qR|^\alpha},\qquad
C(Q)=1+\lambda\,e^{-|QR|^\alpha},\qquad 0<\alpha\le2,
\end{equation}
where $\hbar=1$ units have been utilized. The case $\alpha=2$ corresponds to a Gaussian distribution and $\alpha=1$ to a Cauchy distribution. For $\alpha<2$ the source has a power-law tail,
\begin{equation}
S(\mathbf r)\propto\frac{R^\alpha}{r^{3+\alpha}}\qquad (r\gg R),
\end{equation}
with the corresponding $r^{-(d+\alpha)}$ behaviour in $d$ dimensions. The source parameters $\alpha$, $R$, and $\lambda$ may all depend on $K$, and this dependence is often investigated as a function of pair transverse mass $m_T=\sqrt{K_T^2+m^2}$, where $K_T$ is the transverse component of $K$ (in $c=1$ units).

\subsection{Measurements}
L\'evy-type analyses of two-pion correlations in high-energy heavy-ion collisions have been carried out by PHENIX and STAR in $\sqrt{s_{NN}}=200$\,GeV Au+Au collisions~\cite{PHENIX:2017ino,PHENIX:2024vjp,Kovacs:2023qax,Mukherjee:2023hrz,Kincses:2024sin}, by NA61/SHINE at the CERN SPS~\cite{NA61SHINE:2023qzr}, and by CMS at the LHC~\cite{CMS:2023xyd}, as also summarized in Ref.~\cite{Csanad:2024hva}. The main features of these experimental results are as follows.

\begin{itemize}
\item The index $\alpha$ lies clearly below the Gaussian value, typically between 1 and 2.
\item The L\'evy scale $R(m_T)$ decreases with $m_T$ such that $R^{-2}$ is approximately linear in $m_T$, as expected from hydrodynamics.
\item The intercept $\lambda(m_T)$ shows a characteristic decrease at low $m_T$.
\item Early analyses found $\alpha$ approximately independent of $m_T$~\cite{PHENIX:2017ino}. More recent analyses indicate that $\alpha$ may increase with $m_T$~\cite{PHENIX:2024vjp}. This finding is also in line with simulations~\cite{Porfy:2025sev,Huang:2025edi,Kincses:2025izu}.
\item Scaling with $m_T$ across particle species is observed in data~\cite{Kovacs:2023qax,Mukherjee:2023hrz}, and partly in simulations as well~\cite{Kincses:2026tfa}.
\end{itemize}

Proposed explanations of $\alpha<2$ include anomalous diffusion in the hadronic phase~\cite{Csanad:2007fr}, critical phenomena (relating $\alpha$ to the critical exponent $\eta$)~\cite{Csorgo:2003uv}, and L\'evy walk as a general mechanism including scattering and resonance decays~\cite{Kincses:2024lnv}. Here we ask whether hydrodynamics alone can produce the observed features.

%=====================================================================
\section{Hydrodynamic and freeze-out framework}
\label{sec:framework}
%=====================================================================

\subsection{Perfect-fluid hydrodynamics}
We use $c=\hbar=k_B=1$, metric $g_{\mu\nu}={\rm diag}(1,-1,-1,-1)$, $u^\mu u_\mu=1$, the comoving derivative $D\equiv u^\mu\partial_\mu$, and the projector $\Delta^{\mu\nu}=g^{\mu\nu}-u^\mu u^\nu$. The fluid is described by
\begin{equation}
T^{\mu\nu}=(\varepsilon+p)\,u^\mu u^\nu-p\,g^{\mu\nu},\qquad \partial_\mu T^{\mu\nu}=0,\qquad \partial_\mu(nu^\mu)=0,
\end{equation}
together with the (constant) equation of state (EoS)
\begin{equation}
\varepsilon=\kappa p,\qquad p=nT,\qquad c_s^2=1/\kappa .\label{eq:eos}
\end{equation}
Projecting the energy--momentum equation onto $u_\nu$ and onto $\Delta^{\mu}{}_{\nu}$ gives
\begin{align}
&\partial_\mu\!\left(p^{\kappa/(\kappa+1)}u^\mu\right)=0, \label{eq:energy}\\
&(\kappa+1)\,p\,Du^\mu=\Delta^{\mu\nu}\partial_\nu p .\label{eq:euler}
\end{align}

\subsection{Freeze-out, spectra and correlations}
At freeze-out on a hypersurface $\Sigma$ we use the Cooper--Frye prescription~\cite{Cooper:1974mv} with a Boltzmann distribution and the fugacity expressed via the local density:
\begin{align}
S(x,K)\,\dd^4x&=\frac{g}{(2\pi)^3}\,K^\mu\dd\Sigma_\mu(x)\,\frac{n(x)}{\Teq(T(x))}\,\exp\!\left[-\frac{K\cdot u(x)}{T(x)}\right],\quad\text{with}\label{eq:CF}\\
\Teq(T)&=\frac{g\,m^2T}{2\pi^2}K_2\!\left(\frac mT\right).\label{eq:neq_fugacity}
\end{align}
The divisor $\Teq(T)$ is essential for the discussion in Sec.~\ref{sec:hubble}. The invariant single-particle spectrum is $E\,\dd N/\dd^3K=\int S(x,K)\,\dd^4x$, and the correlation function follows from Eq.~\eqref{eq:CQ}.

\subsection{Gaussian homogeneity lengths and $m_T$ scaling}
In a Gaussian (saddle-point) approximation the HBT radii are the homogeneity lengths of $S(x,K)$ around its maximum $\bar x(K)$:
\begin{equation}
R_{ij}^{-2}(K)=-\partial_i\partial_j\ln S(x,K)\big|_{x=\bar x(K)} .\label{eq:homog}
\end{equation}
For a flow that is linear near the saddle point, $u(r)\simeq Hr$, with a Gaussian geometric profile of width $R_{\rm geom}$ and temperature $T$, this gives the classic result~\cite{Csorgo:1995bi}
\begin{equation}
R^{-2}=R_{\rm geom}^{-2}+\frac{m_T}{T}H^2 .\label{eq:CsL}
\end{equation}
Similarly, boost-invariant longitudinal flow gives for the source size in the corresponding direction the expression $R^2\simeq\tau_f^2T/m_T$~\cite{Makhlin:1987gm}. These relations underlie the observed approximate linearity of $R^{-2}$ in $m_T$.

%=====================================================================
\section{Hubble-type solutions: successes, problems, and a way out}
\label{sec:hubble}
%=====================================================================

\subsection{The accelerationless solution class}
A broad class of 1+3 dimensional exact solutions is built on a comoving scaling variable~\cite{Csorgo:2003ry,Csanad:2009wc,Csanad:2014dpa}. Let $\bar s=\mathcal F(r_x/t,r_y/t,r_z/t)$ with $u^\mu\partial_\mu\bar s=0$ and $\mathcal F$ being an arbitrary partially differentiable function, and let $\nu(\bar s)>0$ also be differentiable, with $\nu(0)=1$. Then
\begin{equation}
u^\mu=\frac{x^\mu}{\tau},\quad
n=n_0\left(\frac{\tau_0}{\tau}\right)^3\nu(\bar s),\quad
p=p_0\left(\frac{\tau_0}{\tau}\right)^{3+3/\kappa},\quad
T=T_0\left(\frac{\tau_0}{\tau}\right)^{3/\kappa}\frac{1}{\nu(\bar s)},\label{eq:hubbleclass}
\end{equation}
with $\tau=\sqrt{t^2-r^2}$ and $p_0=n_0T_0$, solves the hydrodynamic equations of Sec.~\ref{sec:framework}. All members of this class of solutions are accelerationless: $Du^\mu=0$. Freeze-out at constant proper time $\tau=\tau_0$ is a natural choice for these solutions.

\subsection{Successes}
With an ellipsoidal scaling variable, these solutions reproduce measured single-particle spectra, elliptic flow and HBT radii~\cite{Csanad:2009wc}. Multipole generalisations describe higher-order flow harmonics~\cite{Csanad:2014dpa}, and the closely related Buda--Lund parametrisation has been used extensively in phenomenology~\cite{Csorgo:1995bi,Csanad:2003qa}.

\subsection{The L\'evy member and its apparent problem}
A L\'evy-type member of the class is obtained with~\cite{Szentkuti:2026buv}
\begin{equation}
\bar s=\frac{r_x^2+\epsilon_0^2\tau^2}{v_x^2\tau^2}+\frac{r_y^2+\epsilon_0^2\tau^2}{v_y^2\tau^2}+\frac{r_z^2+\epsilon_0^2\tau^2}{v_z^2\tau^2},\qquad
\nu(\bar s)=\left(\frac{s_0}{\bar s}\right)^{\beta},\label{eq:hubblelevy}
\end{equation}
where $v_{x,y,z}$ are expansion rates, $\epsilon_0$ is a dimensionless regulator, and $s_0$, $\beta$ are constants. At large distance $n\propto r^{-2\beta}$, a power law.

Because $Du^\mu=0$, Euler's equation~\eqref{eq:euler} forces the pressure to be uniform on $\tau=$\,const. Hence $T\propto1/n\propto r^{2\beta}$ \emph{increases} outward. Taken at face value this seems unphysical, and it was long considered the main obstacle to L\'evy-type Hubble solutions.

\subsection{Resolution through the fugacity}
In the Cooper--Frye formula~\eqref{eq:CF} the density enters through the fugacity $n/\Teq(T)$, not through $n$ alone. In the hot outer region, $T\gg m$, one has $\Teq\simeq gT^3/\pi^2$, so the emission from the hot tail is strongly suppressed. In Ref.~\cite{Szentkuti:2026buv} it was shown that, with the full $K_2$-dependent fugacity, this solution yields exponential pion spectra, power-law tailed sources, and L\'evy exponents and source sizes in the measured range. Two shortcomings remain. First, the effective temperatures of the pion spectra are about 0.3--0.5\,GeV in the parameter range studied there, above the measured value of about 0.2\,GeV~\cite{PHENIX:2003iij}. Second, the radii do not follow the hydrodynamic $m_T$ scaling: only one of the three radii decreases with $m_T$, while their average stays nearly constant.
 
An asymptotic estimate illustrates the mechanism. With $s_0=3\epsilon_0^2/v^2$, i.e.\ $\nu=1$ at the centre, where $\bar s$ takes its minimum $3\epsilon_0^2/v^2$, the spherically symmetric case ($v_x=v_y=v_z=v$) has $\nu=(1+r^2/R^2)^{-\beta}$ with $R^2=3\epsilon_0^2\tau_0^2$, while $n\propto\nu$ and $T=T_0/\nu$ for $\tau=\tau_0$. For $T\gg m$ the fugacity divisor becomes $\Teq\simeq gT^3/\pi^2\propto\nu^{-3}$, hence $n/\Teq\propto\nu^4$. For $\tau=\tau_0$ we also have $K^\mu\dd\Sigma_\mu=(E-\mathbf k\cdot\mathbf r/t)\,\dd^3r$, which is bounded, and $K\cdot u\simeq(r/\tau_0)(E-\mathbf k\cdot\hat{\mathbf r})$ for $r\gg\tau_0$, where $\hat{\mathbf r}={\mathbf r}/r$. Therefore the $r\to \infty$ asymptotic behavior can be written as
\begin{equation}
S\propto\frac{n}{\Teq(T)}\,e^{-K\cdot u/T}\propto\nu^4\,\exp\!\left[-c(\hat{\mathbf r},K)\,r^{1-2\beta}\right]\propto r^{-8\beta}\qquad(\beta>1/2,\ T\gg m),\label{eq:hubbletail}
\end{equation}
because the exponent vanishes for $\beta>1/2$. For $\beta<1/2$ the tail is a stretched exponential, and $\beta=1/2$ gives $r^{-4}$ with an angle-dependent prefactor. A direct numerical evaluation of the source confirms the asymptotic slope $-8\beta$ (e.g.\ $-6.74$ and $-6.58$ for $\beta=0.85$ and $m_T=0.2$ and 0.8\,GeV at $r=100$--$300$\,fm, in the side direction, for pions with $\tau_0=8$\,fm, $R=5$\,fm and $T_0=140$\,MeV).

Whether this tail is L\'evy-stable in the strict, asymptotic sense follows from a comparison with Sec.~\ref{sec:exp}: a three-dimensional L\'evy-stable source of index $\alpha$ falls off as $r^{-(3+\alpha)}$. Equating the two exponents, $3+\alpha_\infty=8\beta$, gives
\begin{equation}
\alpha_\infty=8\beta-3 .\label{eq:alphainfH}
\end{equation}
The stable range $0<\alpha_\infty<2$ thus corresponds to $3/8<\beta<5/8$. Combined with the condition $\beta\ge1/2$ for a power-law tail, the Hubble-type source is asymptotically L\'evy-stable for
\begin{equation}
\frac12\le\beta<\frac58,\qquad\text{i.e.}\qquad 1\le\alpha_\infty<2 .\label{eq:betawindowH}
\end{equation}
For larger $\beta$ the tail is still a power law, but its second moment is finite; for example, $\beta=0.85$ gives $\alpha_\infty=3.8$. Such a source is L\'evy-like only over a finite range, which is, however, the range probed by L\'evy fits (Sec.~\ref{sec:finiterange}).

The L\'evy-stable window has a price, which is the particle number; we derive it now. On $\tau=\tau_0$ the number of particles is $N=\int n\,u^\mu\dd\Sigma_\mu$. The hypersurface element is $\dd\Sigma_\mu=(1,-\mathbf r/t)\,\dd^3r$, as used above for $K^\mu\dd\Sigma_\mu$, and $u^\mu=x^\mu/\tau_0$, so
\begin{equation}
u^\mu\dd\Sigma_\mu=\frac{1}{\tau_0}\left(t-\frac{r^2}{t}\right)\dd^3r=\frac{\tau_0}{t}\,\dd^3r .\label{eq:udSigma}
\end{equation}
In the outer region $t\simeq r$ and $n\propto\nu\propto r^{-2\beta}$, hence
\begin{equation}
N\propto\int^\infty r^{-2\beta}\,\frac{\tau_0}{r}\,r^2\,\dd r=\tau_0\int^\infty r^{1-2\beta}\,\dd r ,\label{eq:NHubble}
\end{equation}
which converges only for $\beta>1$. A finite particle number therefore requires $\beta>1$, i.e.\ $\alpha_\infty=8\beta-3>5$. In the whole L\'evy-stable window~\eqref{eq:betawindowH}, and also for $\beta=0.85$, the particle number is infinite, although the source at any fixed $K$ is normalisable ($\int\dd^3r\,r^{-8\beta}$ converges for $\beta>3/8$). The two statements are compatible because the particles emitted from the hot tail have large momenta: the divergence resides at high momentum, as the spectrum derived below confirms.
 
The asymptotic slope is, however, reached only slowly. For pions, $T_0\approx m$, so $T=T_0/\nu\ge T_0\approx m$ everywhere, and the condition $T\gg m$, i.e.\ $r\gg R\,(m/T_0)^{1/(2\beta)}$, reduces to $r\gg R$. The range probed by HBT measurements lies mostly at a few $R$, where $\nu^4$ has not yet reached its power law: $\dd\ln\nu^4/\dd\ln r=-8\beta\,r^2/(r^2+R^2)$. The local slope is therefore shallower than $-8\beta$. For $\beta=0.85$, $R=5$\,fm, $\tau_0=8$\,fm, $T_0=140$\,MeV and pions with $m_T=0.2$\,GeV it is $-5.0$ at $r=10$\,fm and $-6.2$ at $r=20$\,fm.
 
The hot, ultra-relativistic tail also shapes the spectrum: at high $m_T$ it produces a power law. The estimate of $K\cdot u$ used above fails in the forward direction, $\hat{\mathbf r}\parallel\mathbf k$, and this direction dominates. We therefore take $E\simeq|\mathbf k|\gg m$ and $r\gg\tau_0$, where $e^{\eta}\simeq2r/\tau_0$, and expand in the angle $\theta$ between $\hat{\mathbf r}$ and $\mathbf k$. For $\theta\ll1$, $\cosh\eta-\sinh\eta=e^{-\eta}\simeq\tau_0/(2r)$, $\sinh\eta=r/\tau_0$ and $1-\tanh\eta\simeq\tau_0^2/(2r^2)$ give
\begin{equation}
K\cdot u\simeq E\left[\frac{\tau_0}{2r}+\frac{r}{\tau_0}\frac{\theta^2}{2}\right],\qquad
E-\frac{\mathbf k\cdot\mathbf r}{t}\simeq E\left[\frac{\tau_0^2}{2r^2}+\frac{\theta^2}{2}\right].
\end{equation}
With $\dd\Omega\simeq\pi\,\dd(\theta^2)$, the integral over the cone is Gaussian:
\begin{equation}
\int\dd\Omega\left(E-\frac{\mathbf k\cdot\mathbf r}{t}\right)e^{-K\cdot u/T}\simeq\frac{2\pi\tau_0^2T^2}{E\,r^2}\,(1+X)\,e^{-X},\qquad X=\frac{E\tau_0}{2rT}.
\end{equation}
In the tail, $n/\Teq\propto\nu^4\propto r^{-8\beta}$ and $T\propto r^{2\beta}$. Hence $X=(r_*/r)^{1+2\beta}$ with $r_*^{1+2\beta}=E\tau_0R^{2\beta}/(2T_0)$, and with $r=r_*y$
\begin{equation}
E\frac{\dd N}{\dd^3K}\propto\frac1E\int\dd r\,r^{-4\beta}(1+X)e^{-X}
=\frac{r_*^{1-4\beta}}{E}\int_0^\infty\dd y\,y^{-4\beta}\left(1+y^{-1-2\beta}\right)e^{-y^{-1-2\beta}}\propto E^{-1+\frac{1-4\beta}{1+2\beta}}=E^{-\frac{6\beta}{1+2\beta}} ,
\end{equation}
where the $y$ integral converges for $\beta>1/4$, and we used $r_*\propto E^{1/(1+2\beta)}$. At $y=0$, $E=m_T$, so for $r_*\gg\tau_0,R$ the spectrum falls as $m_T^{-6\beta/(1+2\beta)}$: as $m_T^{-1.57}$, $m_T^{-1.75}$ and $m_T^{-1.89}$ for $\beta=0.55$, 0.7 and 0.85. (For a generic direction, the estimate used for Eq.~\eqref{eq:hubbletail} gives a contribution $\propto[E(1-\cos\theta)]^{1-q}$ with $q=(8\beta-3)/(2\beta-1)$. For $q>2$ its angular integral is dominated by $\theta\to0$, where that estimate fails; this is why the forward cone dominates.) A direct numerical evaluation ($\tau_0=8$\,fm, $R=5$\,fm, $T_0=140$\,MeV, pions) gives local slopes of $-1.56$ to $-1.57$, $-1.74$ to $-1.75$ and $-1.87$ to $-1.89$ at $m_T=10$--$300$\,GeV. This power law is much harder than measured pion spectra, and it sets in early: for $\beta=0.85$ the inverse slope already grows from 0.54\,GeV at $m_T=0.7$\,GeV to 1.6\,GeV at 3\,GeV. It also carries the infinite particle number found above. For a spherically symmetric source $N=\int(\dd^3K/E)\,E\,\dd N/\dd^3K\propto\int^\infty E\,\dd E\,E^{-6\beta/(1+2\beta)}$, which converges only if $6\beta/(1+2\beta)>2$, i.e.\ $6\beta>2+4\beta$, i.e.\ $\beta>1$: the same condition as from Eq.~\eqref{eq:NHubble}. As also discussed later in Sec.~\ref{sec:general}, this is the ``fast-tail'' route to power-law tailed sources: the flow becomes ultra-relativistic, so a temperature growing at least linearly with $r$ is not a pathology but a requirement.

%=====================================================================
\section{Accelerating exact solutions: a brief history}
\label{sec:history}
%=====================================================================

The first exact solution of relativistic hydrodynamics relevant to high-energy collisions, Landau's 1+1 dimensional accelerating flow~\cite{Landau:1953gs}, was obtained with the hodograph method developed further by Khalatnikov~\cite{Khalatnikov:1954aa}. Bjorken's boost-invariant flow~\cite{Bjorken:1982qr} is accelerationless and became the standard reference for the longitudinal dynamics.

Interpolations between the two were obtained by Bialas, Janik and Peschanski~\cite{Bialas:2007iu} using a potential (Khalatnikov) formulation, with further developments in Ref.~\cite{Beuf:2008vd}. Csörgő and collaborators found a family of accelerating, explicit 1+1 dimensional solutions and computed their observables~\cite{Csorgo:2006ax,Nagy:2007xn}. More recent work extended these families~\cite{Csorgo:2018pxh,Csorgo:2018fbz}.

Conformal symmetry provides another route. Gubser constructed a boost-invariant solution with transverse expansion by mapping a static fluid in de Sitter space to Minkowski space~\cite{Gubser:2010ze,Gubser:2010ui}. It was later generalised to viscous fluids~\cite{Marrochio:2013wla} and to non-boost-invariant flows~\cite{Hatta:2014gqa,Hatta:2014gga}. Self-similar relativistic flows are also well studied in astrophysics, e.g.\ blast waves~\cite{Blandford:1976uq}.

Despite this progress, a solution that is relativistic, accelerating, genuinely 1+3 dimensional, and gives realistic observables is still missing.

%=====================================================================
\section{General results}
\label{sec:general}
%=====================================================================

\subsection{Dressing theorem}
Equations~\eqref{eq:energy}--\eqref{eq:euler} involve only $p$ and $u^\mu$, so flow and pressure form a closed system for any $\kappa$. Define
\begin{equation}
w\equiv p^{1/(\kappa+1)} .
\end{equation}

\noindent\textbf{Theorem (dressing).} Let $(p,u^\mu)$ solve Eqs.~\eqref{eq:energy}--\eqref{eq:euler}, and let $\nu>0$ satisfy $u^\mu\partial_\mu\nu=0$. Then
\begin{equation}
n=w^\kappa\,\nu,\qquad T=\frac{w}{\nu}\label{eq:dressing}
\end{equation}
solve the full system including $\partial_\mu(nu^\mu)=0$, with $nT=p$.

\noindent\emph{Proof.} Since $nT=w^{\kappa+1}=p$, the EoS holds. Furthermore $\partial_\mu(nu^\mu)=\nu\,\partial_\mu(w^\kappa u^\mu)+w^\kappa D\nu=0$ by Eq.~\eqref{eq:energy} and $D\nu=0$. $\square$

The theorem is elementary, and special cases of it are well known. In the Hubble-type class~\eqref{eq:hubbleclass}, the arbitrary scaling function $\nu(\bar s)$ plays exactly this role~\cite{Csorgo:2003ry,Csanad:2009wc}, and the accelerating 1+1 dimensional solutions of Refs.~\cite{Csorgo:2006ax,Nagy:2007xn} contain an analogous arbitrary scaling function. What we use below is that the freedom exists for every flow: any solution of Eqs.~\eqref{eq:energy}--\eqref{eq:euler} can be dressed.

The function $\nu$ encodes the entropy per particle $\sigma$ as $\nu \propto e^{-\sigma/(\kappa+1)}$, which is conserved along streamlines, as both $u^\mu \partial_\mu \sigma = 0$ and $u^\mu \partial_\mu \nu = 0$. In other words, both $\sigma$ and $\nu$ are constant following a fluid element. Equation~\eqref{eq:hubbleclass} is the special case of Hubble flow, and the convention~\eqref{eq:dressing} matches it ($n\propto\nu$, $T\propto1/\nu$). In non-pathologic cases where every streamline crosses the freeze-out hypersurface $\Sigma$ exactly once, $\nu$ can be chosen by imposing the boundary condition

\begin{align}
\left.\nu\right|_\Sigma=\frac{p^{1/(\kappa+1)}}{T_f}
\end{align}

and then extending $\nu$ along streamlines according to $u^\mu\partial_\mu\nu=0$. This yields $T=T_f$ on $\Sigma$, which we call the \emph{isothermal dressing}. Then $n=p/T_f$ on $\Sigma$, and the source shape is determined by $p$ and $u$ on $\Sigma$ alone.

Fixing instead the entropy per particle on $\Sigma$, $\sigma|_\Sigma=\sigma_f$, gives the \emph{isentropic dressing}. Since $\sigma$ is conserved along streamlines, it is then the same everywhere: $\nu=$\,const, hence $n\propto w^\kappa = p^{\kappa/(\kappa+1)}$ and $T\propto w = p^{1/(\kappa+1)}$.

\subsection{Potential formulation}
For irrotational flows we write
\begin{equation}
w\,u_\mu=\partial_\mu\Phi, \qquad w^2=\partial_\mu\Phi\,\partial^\mu\Phi .
\label{eq:potential}
\end{equation}
The second relation follows from $u_\mu u^\mu=1$. Moreover,
\begin{align}
D(wu_\mu) &=u^\nu\partial_\nu\partial_\mu\Phi  =u^\nu\partial_\mu(wu_\nu)  =\partial_\mu w,\;\textnormal{hence}\\
wDu_\mu&=(\partial_\mu-u_\mu D)w =\Delta_{\mu\nu}\partial^\nu w .
\end{align}
Since $p=w^{\kappa+1}$, this is precisely Euler's equation~\eqref{eq:euler}; thus Euler's equation is satisfied identically by the potential representation.

The remaining energy equation~\eqref{eq:energy} becomes
\begin{equation}
\partial_\mu(w^\kappa u^\mu)=0.
\end{equation}
Using $w u^\mu=\partial^\mu\Phi$, it can therefore be written as the single nonlinear field equation
\begin{equation}
\partial_\mu\!\left(w^{\kappa-1}\partial^\mu\Phi\right)=0,
\qquad
w^2=\partial_\mu\Phi\,\partial^\mu\Phi .
\label{eq:PhiPDE}
\end{equation}
This is the relativistic Khalatnikov--Bialas--Janik--Peschanski formulation~\cite{Khalatnikov:1954aa,Bialas:2007iu}. Equation~\eqref{eq:PhiPDE} is the Euler--Lagrange equation of the Lagrangian $L=(\partial_\mu\Phi\,\partial^\mu\Phi)^{(\kappa+1)/2}$, which equals the pressure, since $\partial L/\partial(\partial_\mu\Phi)=(\kappa+1)\,w^{\kappa-1}\partial^\mu\Phi$. In this scalar-field form, potential flows of a fluid with a linear equation of state were studied for an arbitrary $\kappa$ in Ref.~\cite{Borshch:2007uf}. What we add here is the dressing theorem, through which the formulation does not assume a uniform entropy per particle.

\subsection{Flow mismatch and power-law tails}\label{sec:mismatch}
For a pair momentum $K$, Eq.~\eqref{eq:CF} contains
\begin{align}
\exp\!\left[-\frac{K\cdot u-m}{T}\right]\,
\frac{n\,e^{m/T}}{\Teq(T)}.
\end{align}

One can then introduce the $K^\mu=m_T(\cosh Y_K,\sinh Y_K,\mathbf{K}_T/m_T)$ pair momentum parameterisation, while the local flow four-velocity may be parametrised in terms of its longitudinal rapidity $Y_u$ (and, more generally, transverse flow). Denoting by $\Delta Y$ the rapidity difference between the local flow and the pair velocity, one has, for purely longitudinal motion, $\Delta Y=Y_u-Y_K$ and $K\cdot u=m_T\cosh\Delta Y$, hence
$K\cdot u-m_T=m_T(\cosh\Delta Y-1)\geq 0$.

For the particularly simple case of a comoving pair at rest in the transverse plane, this reduces to the corresponding expression with $m$ in place of $m_T$. Thus, when the flow becomes ultra-relativistic in the outer regions, $|\Delta Y|$ grows with the coordinate rapidity and the Boltzmann factor becomes exponentially suppressed unless the temperature grows sufficiently rapidly.

Two simple routes to a power-law tailed source at fixed $K$ therefore remain:
\begin{enumerate}
\item \emph{Fast tail}: $T$ grows at least linearly with the radial coordinate. This is the Hubble route of Sec.~\ref{sec:hubble}, in which the growth of the temperature compensates the increasing $K\cdot u$, while the fugacity factor keeps the emission from the hot region under control.

\item \emph{Slow tail}: the Lorentz factor of the asymptotic flow stays bounded, so that $K\cdot u$ stays bounded, and the freeze-out temperature tends to a constant. On this isothermal freeze-out hypersurface, the Boltzmann equation of state gives $n\propto p$; hence a power-law tailed pressure profile implies a similar density profile as well. The Boltzmann factor then approaches a constant at large radius, so the source inherits this power-law tail.
\end{enumerate}
In the slow-tail scenario, the asymptotic emission is suppressed relative to the comoving core by a factor of order $e^{-(m_T-m)/T}$. This factor decreases with $m_T$, so the relative weight of the power-law tail decreases with $m_T$, and the effective L\'evy exponent $\alpha$ increases towards 2. This provides a natural mechanism for an $m_T$-dependent $\alpha$, with the same trend as indicated by recent measurements (Sec.~\ref{sec:exp}).

\subsection{The pressure bounds the source}\label{sec:bound}
Before discussing the two routes further, we derive a bound that both of them have to respect. It follows from the equation of state and the Cooper--Frye formula alone, independently of the dressing.

We first rewrite the fugacity. With $p=nT$, and with the pressure of an ideal Boltzmann gas of the emitted particles, $p_{\rm eq}(T)=T\,\Teq(T)$, the fugacity in Eq.~\eqref{eq:CF} becomes
\begin{equation}
\frac{n}{\Teq(T)}=\frac{p/T}{p_{\rm eq}(T)/T}=\frac{p}{p_{\rm eq}(T)} .\label{eq:fugp}
\end{equation}
Next we bound the other factors of Eq.~\eqref{eq:CF} for freeze-out at $\tau=\tau_f$:
\begin{itemize}
\item $p_{\rm eq}(T)=g\,m^2T^2K_2(m/T)/(2\pi^2)$ increases monotonically with $T$.
\item In the local rest frame of the fluid, $K\cdot u$ is the energy of the particle, so $K\cdot u\ge m>0$, and the Boltzmann factor $e^{-K\cdot u/T}$ is at most 1.
\item The flux factor per unit volume, $K^\mu\dd\Sigma_\mu/\dd^3r=E-\mathbf k\cdot\mathbf r/t$, is bounded by $E+|\mathbf k|$, as in Sec.~\ref{sec:hubble}.
\end{itemize}
Hence, if the freeze-out temperature is bounded from below, $T\ge T_{\min}>0$, the source satisfies
\begin{equation}
S(x,K)\le\frac{g\,(E+|\mathbf k|)}{(2\pi)^3\,p_{\rm eq}(T_{\min})}\;p(x) .\label{eq:bound}
\end{equation}
The source tail can therefore never be heavier than the pressure tail on the freeze-out hypersurface. In particular, a L\'evy-stable source, $S\propto r^{-(3+\alpha)}$ with $\alpha<2$, requires a pressure that decays at most as $r^{-(3+\alpha)}$, and thus more slowly than $r^{-5}$. A temperature that drops to zero in the tail does not evade the bound either. For $T\ll m$, $p_{\rm eq}\propto T^{5/2}e^{-m/T}$. Where the flow becomes static, $K\cdot u\to m_T$, and the source behaves as $p\,T^{-5/2}e^{-(m_T-m)/T}$. As a function of $T$, the factor $T^{-5/2}e^{-(m_T-m)/T}$ is maximal at $T=2(m_T-m)/5$, so $p\,T^{-5/2}e^{-(m_T-m)/T}\le[5/(2e(m_T-m))]^{5/2}\,p$. For $m_T>m$ the source is therefore still bounded by the pressure, with a constant that diverges as $K_T\to0$.

The bound shows how the two routes differ. In the slow-tail route $T$ tends to a constant, so $p_{\rm eq}(T)$ does too, and the bound is saturated up to a constant factor: the source tail is the pressure tail. This is why the slow-tail route needs a L\'evy-like pressure, which is the subject of the next subsection. In the fast-tail route $T$ grows outwards, $p_{\rm eq}(T)$ grows with it, and the source falls off faster than the pressure. The pressure must then be flatter than the desired source tail. Hubble-type solutions satisfy this trivially, since their pressure is uniform on $\tau=$\,const. For \lPF{} and for the dressed Gubser flow the bound is applied in Secs.~\ref{sec:obsSolI} and~\ref{sec:gubsertail}, respectively.

The fast-tail route has a second limitation, which concerns the particle number. We give the general argument here, for freeze-out at $\tau=\tau_f$ and an outer flow that differs from Hubble flow by a constant rapidity $\zeta$; the explicit exponents follow in Secs.~\ref{sec:hubble} and~\ref{sec:obsSolI}. The argument compares two densities per unit volume $\dd^3r$. The first is that of the particle number, $n\,u^\mu\dd\Sigma_\mu/\dd^3r$. In the outer region $u^\mu\dd\Sigma_\mu=\cosh\zeta\,(\tau_f/t)\,\dd^3r\propto\dd^3r/r$ [Eq.~\eqref{eq:udSigma} for Hubble flow; the rapidity offset contributes the factor $\cosh\zeta$], so this density is $\propto n/r=p/(Tr)$, and the particle number is finite only if $p/(Tr)$ falls off faster than $r^{-3}$. The second is the source at fixed $K$. Once $T\gg m$ and the Boltzmann factor has become constant, it is $\propto p/p_{\rm eq}(T)\propto p/T^4$, which we write as $[p/(Tr)]\,(r/T^3)$. A power-law tail requires $T$ to grow at least linearly with $r$, so the factor $r/T^3$ falls off at least as $r^{-2}$. With a finite particle number the source therefore falls off faster than $r^{-5}$, i.e.\ faster than any L\'evy-stable tail. In the fast-tail route, a L\'evy-stable tail thus requires an infinite particle number. The slow-tail route has no such restriction: there the flow is slow, $u^\mu\dd\Sigma_\mu\propto\dd^3r$, the particle density is proportional to the source itself, and a tail $\propto r^{-(3+\alpha)}$ is integrable for every $\alpha>0$.

The two routes are sufficient, not exhaustive. Since the bound only requires $T\ge T_{\min}$, a slow flow with an outward-growing temperature also gives a power-law tail, which the fugacity makes steeper than the pressure tail. In every slow-flow case, however, the pressure must not fall off faster than a power law, and a power-law decay of the pressure requires the $1/r$ acceleration.

\subsection{The $1/r$-acceleration criterion}
In the slow-tail route with isothermal dressing, a L\'evy-like density requires a L\'evy-like pressure. Euler's equation~\eqref{eq:euler} relates the pressure profile to the proper acceleration,
\begin{equation}
a^\mu\equiv u^\nu\partial_\nu u^\mu = \frac{\Delta^{\mu\nu}\partial_\nu\ln p}{\kappa+1} .\label{eq:accEuler}
\end{equation}
We now evaluate this relation in the outer region of a radial flow with velocity $v$ and Lorentz factor $\gamma$. The proper acceleration is orthogonal to $u^\mu=\gamma(1,v)$, so in the $(t,r)$ plane it is parallel to the outward unit vector $n^\mu=\gamma(v,1)$. We write $a^\mu=a\,n^\mu$, where $a$ is counted positive outwards. Contracting Eq.~\eqref{eq:accEuler} with $n_\mu$, and using $n_\mu n^\mu=-1$ together with $n_\mu\Delta^{\mu\nu}=n^\nu$, gives
\begin{equation}
(\kappa+1)\,a=-n^\nu\partial_\nu\ln p=-\gamma\left(\partial_r\ln p+v\,\partial_t\ln p\right) .\label{eq:aslow}
\end{equation}
The first term is the pressure gradient. The second term is the change of the pressure seen by a fluid element that moves outwards. For a power-law pressure, $p\propto r^{-(3+\alpha)}$, the first term equals $(3+\alpha)/r$. In a quasi-static tail, where $\gamma\simeq1$ and $|v\,\partial_t\ln p|\ll1/r$, the second term can be neglected, and
\begin{equation}
p\propto r^{-(3+\alpha)}\;\Longleftrightarrow\;a\simeq\frac{3+\alpha}{(\kappa+1)\,r}\qquad\text{(quasi-static tail)} .\label{eq:criterion}
\end{equation}
If the tail instead expands self-similarly, with $v\propto t/r$ and $\partial_t\ln p\propto1/t$, then the second term in Eq.~\eqref{eq:aslow} also falls off as $1/r$. It changes the coefficient of the $1/r$ law, but not the law itself; the dressed Gubser flow of Sec.~\ref{sec:solII} is an example. In both cases a Gaussian pressure corresponds to an acceleration linear in $r$ (a harmonic potential), whereas a L\'evy-like pressure corresponds to an acceleration that falls off as $1/r$ (a logarithmic potential). Accelerating solutions are therefore not optional for the slow-tail route; they are its defining feature. Two explicit realisations of the $1/r$ acceleration are given in Appendix~\ref{app:companion}: a self-similar family describing the evolution of power-law initial data, whose outer branch has $v\simeq a_p\,t/[(\kappa+1)\,r]$, where $a_p$ is the exponent of the initial pressure, $p_0\propto r^{-a_p}$, and an exact spherical Rindler atmosphere with proper acceleration $1/\rho$, which corresponds to a L\'evy index $\alpha=\kappa-2$.

\subsection{Causality and finite range}\label{sec:finiterange}
Hydrodynamic evolution from a compact initial state keeps the fluid inside the causal envelope $r<R_0+t$. Strict power-law tails therefore require either initial data with infinite extent or an idealised exact solution. Experimentally, L\'evy fits probe a finite range of $Q$ (equivalently of $r$). A ``L\'evy-like'' source in the experimental sense is therefore a statement about a finite range, and we treat it as such numerically.

%=====================================================================
\section{\texorpdfstring{Self-similar $\lambda$ potential flows (\lPF)}{Self-similar lambda potential flows (lambda-PF)}}
\label{sec:solI}
%=====================================================================

This section presents an explicit family of relativistic, accelerating, self-similar potential flows that contains Hubble flow as a special case. It is also the interior of the self-similar flow with power-law initial data of Appendix~\ref{app:companion}. A single parameter, the exponent $\lambda$ of the potential in Eq.~\eqref{eq:ansatz}, controls the acceleration. We therefore call these flows $\lambda$ potential flows, or \lPF{} for short. The name should not be confused with the accelerating 1+1 dimensional solutions of Refs.~\cite{Csorgo:2006ax,Nagy:2007xn}, which are sometimes also called $\lambda$ solutions: there $\lambda=1$ gives Hubble flow, whereas here Hubble flow corresponds to $\lambda=\lH=1-3/\kappa$ (Sec.~\ref{sec:central}). The parameter $\lambda$ is also unrelated to the intercept $\lambda(K)$ of Eq.~\eqref{eq:CQ}. Two of its members are known: Hubble flow, and the $\lambda=0$ member, which is the self-similar solution $\Phi(r/t)$ of Ref.~\cite{Borshch:2007uf} (Sec.~\ref{sec:central}). The ansatz with a general $\lambda$ is mentioned in Ref.~\cite{Borshch:2007uf} as a possible generalisation, but the resulting equation was not derived or solved there. To our knowledge the family has not been analysed elsewhere in the heavy-ion literature. Self-similar relativistic flows have, however, a long history in other contexts, e.g.\ Ref.~\cite{Blandford:1976uq}, and we cannot exclude that equivalent solutions have appeared there in a different parametrisation.

Before turning to the details, we state what \lPF{} contributes, since it does not by itself produce L\'evy-shaped sources. We use it for three purposes. First, it shows how much acceleration a regular self-similar flow expanding from a point can carry (Sec.~\ref{sec:solIaccel}). Second, it shows that this kind of acceleration does not produce L\'evy tails: the pressure on $\tau=$\,const decays at most like about $r^{-1}$, and the acceleration in the outer region does not fall off as $1/r$, so the criterion~\eqref{eq:criterion} is not met. Third, it provides an accelerating background for the fast-tail dressing of Sec.~\ref{sec:hubble}, which generalises the Hubble-type L\'evy source, including its L\'evy-stable tail, to an accelerating core (Sec.~\ref{sec:obsSolI}). Readers mainly interested in L\'evy-like sources may skip to Sec.~\ref{sec:solII}.

\subsection{Ansatz and field equation}
We consider spherically symmetric potentials of the self-similar form
\begin{equation}
\Phi(t,r)=t^\lambda F(s),\qquad s=\frac{r^2}{t^2}\in[0,1) ,\label{eq:ansatz}
\end{equation}
and define
\begin{align}
A&=\lambda F-2sF',\label{eq:Adef}\\
B&=\lambda F+2(1-s)F'=A+2F',\label{eq:Bdef}\\
H^2&=A^2-4sF'^2 .\label{eq:H2def}
\end{align}
Then $w=t^{\lambda-1}H(s)$, $u^0=A/H$, and the radial velocity is
\begin{equation}
v=-\frac{2\sqrt s\,F'}{A} .\label{eq:vF}
\end{equation}
From Eqs.~\eqref{eq:Adef}, \eqref{eq:H2def} and~\eqref{eq:vF}, $1-v^2=H^2/A^2$; physical (subluminal) flows therefore require $H^2>0$, equivalently a timelike $\partial_\mu\Phi$ and a real, positive $w=p^{1/(\kappa+1)}$. Subsequently we may observe that Eq.~\eqref{eq:PhiPDE} reduces to
\begin{equation}
\kappa(\lambda-1)A+4s(s-1)F''-\big[2s(\lambda-2)+6\big]F'-(\kappa-1)\,s\,B\,\frac{(H^2)'}{H^2}=0 .\label{eq:FODE}
\end{equation}
One may verify this reduction symbolically: substituting the ansatz into Eq.~\eqref{eq:PhiPDE} gives Eq.~\eqref{eq:FODE} multiplied by the positive factor $t^{\lambda-\kappa-1}(t^{2\lambda}H^2)^{(\kappa-1)/2}$. Solving for $F''$ gives
\begin{equation}
F''=\frac{[2s(\lambda-2)+6]F'-\kappa(\lambda-1)A+\dfrac{(\kappa-1)sB}{H^2}\left[2A(\lambda-2)F'-4F'^2\right]}{4s\left[(s-1)+(\kappa-1)sB^2/H^2\right]} .\label{eq:Fpp}
\end{equation}

The centre ($s=0$) requires special care, because the coefficient of $F''$ in Eq.~\eqref{eq:FODE} vanishes there, i.e., the denominator in Eq.~\eqref{eq:Fpp} is zero. Eq.~\eqref{eq:FODE} is of second order and thus has two integration constants; we now show that at this singular point one of them is fixed by requiring a physical flow.

We first look for solutions that are smooth at the centre, i.e.\ with $F$, $F'$ and $F''$ finite at $s=0$. Setting $s=0$ in Eq.~\eqref{eq:FODE} removes every term that carries an explicit factor of $s$. With $A(0)=\lambda F(0)$, only
\begin{equation}
\kappa\lambda(\lambda-1)F(0)-6F'(0)=0
\end{equation}
remains. For a smooth solution the initial slope is therefore not free but fixed to $F'(0)=\kappa\lambda(\lambda-1)F(0)/6$, and the higher coefficients of its power series in $s$ follow in the same way (see Sec.~\ref{sec:central} below). This smooth solution, which we call the \emph{regular branch}, contains only one integration constant, $F(0)$.

The second integration constant belongs to a solution that is not smooth at the centre. To identify it, consider a small deviation $\delta F$ from the regular branch with a local power-law behaviour $\delta F\propto s^\mu$. Inserting $F+\delta F$ into Eq.~\eqref{eq:FODE} and keeping the terms linear in $\delta F$, the two terms $-4s\,\delta F''$ and $-6\,\delta F'$ are of order $s^{\mu-1}$, whereas all other terms carry an additional factor of $s$ and are of order $s^\mu$. For $s\to0$ the two leading terms must therefore cancel on their own, which requires
\begin{equation}
-4\mu(\mu-1)-6\mu=-2\mu(2\mu+1)=0 ,
\end{equation}
i.e.\ $\mu=0$ or $\mu=-1/2$. To linear order, the general solution near the centre is therefore
\begin{equation}
F\simeq C_{\rm reg}\,F_{\rm reg}(s)+C\,s^{-1/2},
\end{equation}
where $C_{\rm reg}$ rescales the regular branch ($\mu=0$) and $C$ is the amplitude of a singular component ($\mu=-1/2$). This component, $\propto t/r$, describes a point source at the origin and grows towards the centre. Where it dominates $F$, one finds $v\simeq s^{-1/2}/(\lambda+1)>1$ and $H^2\simeq-4sF'^2<0$, i.e.\ a superluminal flow without a real pressure. Physical solutions therefore require $C=0$. This is the same as requiring smoothness at the centre: a solution with $C\neq0$ does not have a different finite $F'(0)$, its slope diverges. All physical solutions thus lie on the regular branch.

Finally, since Eq.~\eqref{eq:FODE} is homogeneous of degree one in $F$, the remaining constant $F(0)$ only sets the overall normalisation of the pressure, $p\propto|F(0)|^{\kappa+1}$. We take $|F(0)|=1$, with the sign of $F(0)$ equal to that of $\lambda$, so that $u^0=A/H>0$. With this choice, $\lambda$ is the only parameter of the regular branch.

\subsection{Central expansion and the Hubble limit}\label{sec:central}
The regular series $F=F_0+F_1s+F_2s^2+\dots$ has
\begin{equation}
F_1=\frac{\kappa\lambda(\lambda-1)}{6}F_0 ,
\end{equation}
which is equivalent to $F_1=F'(0)$ as by regularity, see above; and $F_2$ is given in Appendix~\ref{app:F2}. Near the centre $v\simeq c\,r/t$ with
\begin{equation}
c=\frac{\kappa(1-\lambda)}{3} .
\end{equation}
Fluid elements follow $\dd\xi/\dd\ln t=v(\xi)-\xi$ with $\xi=r/t$. Hence $c>1$ ($\lambda<\lH$) is outward acceleration and $c<1$ is deceleration, where
\begin{equation}
\lH=1-\frac3\kappa,\qquad F_{\rm H}=(1-s)^{\lH/2}\label{eq:FH}
\end{equation}
is the exact Hubble solution. For $\kappa=3$ one has $\lH=0$, and the Hubble potential is $\Phi=\ln\tau$, which lies outside the power ansatz. It is nevertheless the limit $\lambda\to0$ of the regular branch. Since Eq.~\eqref{eq:PhiPDE} is invariant under $\Phi\to a\Phi+b$, the potential $[\Phi/F(0)-1]/\lambda$, which differs from $\Phi$ only by the positive factor $1/|\lambda|$ and a constant, describes the same flow, and $(\tau^\lambda-1)/\lambda\to\ln\tau$. Indeed, at bounded $\eta$ the regular branch for $\kappa=3$ has $\Phi/F(0)=\tau^\lambda[1+\mathcal O(\lambda^2)]$; at large $\eta$, $G\propto e^{\delta\eta}$ with $\delta\simeq3\lambda^2/2$ (Sec.~\ref{sec:milne} and Eq.~\eqref{eq:lightcone} below). Its velocity deviates from Hubble flow only at first order in $\lambda$. To this order, the first-order form~\eqref{eq:ueq} of the field equation, derived in Sec.~\ref{sec:milne}, becomes $u'=3\lambda-2u\coth\eta$, i.e.\ $(u\sinh^2\eta)'=3\lambda\sinh^2\eta$, so $u=\frac{3\lambda}{2}\,\frac{\sinh\eta\cosh\eta-\eta}{\sinh^2\eta}$. Then $v-r/t=\tanh(\eta-{\rm artanh}\,u)-\tanh\eta\simeq-u/\cosh^2\eta$, whose maximum is $|v-r/t|=0.417\,\lambda$, at $\eta\approx0.71$.

Let us briefly discuss the $\lambda=0$ case as well. Within the ansatz~\eqref{eq:ansatz}, $\lambda=0$ gives scale-invariant potentials $\Phi=F(s)$, and these cannot describe a flow inside the light cone. To see this, note that $\Phi$ is homogeneous of degree $\lambda$ in $x^\mu=(t,\mathbf r)$, so Euler's theorem for homogeneous functions gives
\begin{equation}
x^\mu\partial_\mu\Phi=w\,u\cdot x=\lambda\Phi .
\end{equation}
Inside the light cone both $x^\mu$ and $u^\mu$ are timelike and future-directed, so $u\cdot x>0$. For $\lambda=0$ this requires $w=0$, i.e.\ no flow at all. (For $\lambda\neq0$ the same identity shows why $\Phi$, and hence $F(0)$, must have the sign of $\lambda$.) In terms of Eqs.~\eqref{eq:Adef}--\eqref{eq:H2def}, $\lambda=0$ gives
\begin{equation}
A=-2sF',\qquad B=2(1-s)F',\qquad H^2=-4s(1-s)F'^2 ,
\end{equation}
so $H^2<0$ for $0<s<1$ whenever $F'\neq0$. The field equation~\eqref{eq:FODE} nevertheless has non-trivial solutions for $\lambda=0$. With $(H^2)'/H^2=1/s-1/(1-s)+2F''/F'$, its four terms become $2\kappa sF'$, $4s(s-1)F''$, $-(6-4s)F'$, and $-2(\kappa-1)(1-2s)F'-4(\kappa-1)s(1-s)F''$. Their sum vanishes if
\begin{equation}
\frac{F''}{F'}=\frac{6\kappa s-2\kappa-4}{4\kappa s(1-s)}=-\frac{\kappa+2}{2\kappa s}+\frac{\kappa-1}{\kappa(1-s)} ,
\end{equation}
a first-order equation for $F'$, which integrates to
\begin{equation}
F'=C\,s^{-(\kappa+2)/(2\kappa)}\,|1-s|^{-(\kappa-1)/\kappa}\label{eq:Flam0}
\end{equation}
($F'=C\,s^{-5/6}|1-s|^{-2/3}$ for $\kappa=3$). Inside the light cone this solution has $H^2<0$ and is therefore not a flow. Outside the light cone, $s>1$, the sign of $H^2$ is reversed, and Eq.~\eqref{eq:Flam0} describes a physical flow: the exact spherical Rindler atmosphere of Appendix~\ref{app:companion}. Indeed, with $\rho=\sqrt{r^2-t^2}=t\sqrt{s-1}$, the gradient
\begin{equation}
\partial_\mu\Phi=F'\,\partial_\mu s=\frac{2F'}{t}\left(-s,\sqrt s\right)
\end{equation}
gives $u^\mu\propto(\sqrt s,1)\propto(r,t)$, i.e.\ $u^\mu=(r,t)/\rho$ for $C<0$. Its norm, $w^2=4s(s-1)F'^2/t^2$, becomes with Eq.~\eqref{eq:Flam0}
\begin{equation}
w=\frac{2|C|}{\rho}\left(\frac{\rho}{r}\right)^{2/\kappa},\qquad
p=w^{\kappa+1}\propto\frac{1}{\rho^{\kappa+1}}\left(\frac{\rho}{r}\right)^{2(\kappa+1)/\kappa},
\end{equation}
which is Eq.~\eqref{eq:rindler}. The Rindler atmosphere is thus the $\lambda=0$ member of the self-similar potential flows, continued to the outside of the light cone. In this form, as a potential that depends on $r/t$ only, the solution was obtained in Ref.~\cite{Borshch:2007uf}.

\subsection{Milne form and sonic points}\label{sec:milne}
With $t=\tau\cosh\eta$ and $r=\tau\sinh\eta$ one has $s=\tanh^2\eta$, and, since $t^\lambda=\tau^\lambda\cosh^\lambda\eta$, the ansatz becomes $\Phi=\tau^\lambda G(\eta)$ with $G=F\cosh^\lambda\eta$.

In this form the field equation reduces to a first-order equation. With $\sqrt{-g}\propto\tau^3\sinh^2\eta$, $w=\tau^{\lambda-1}h$ and $h^2=\lambda^2G^2-G'^2$, Eq.~\eqref{eq:PhiPDE} becomes, after the common power of $\tau$ is divided out,
\begin{equation}
\left(\sinh^2\eta\,h^{\kappa-1}G'\right)'=\kappa\lambda(\lambda-\lH)\,\sinh^2\eta\,h^{\kappa-1}G ,
\end{equation}
where $3+\kappa(\lambda-1)=\kappa(\lambda-\lH)$ was used. We introduce $u\equiv G'/(\lambda G)$. By Eq.~\eqref{eq:potential}, $w\cosh\zeta=\partial_\tau\Phi=\lambda\tau^{\lambda-1}G$ and $-w\sinh\zeta=\tau^{-1}\partial_\eta\Phi=\tau^{\lambda-1}G'$, where $\zeta$ is the rapidity of the flow relative to Hubble flow. Hence $u=-\tanh\zeta$, and $|u|$ is the relative velocity $v_{\rm rel}$ of Eq.~\eqref{eq:vrel} below. Expanding the left-hand side gives $\sinh^2\eta\,h^{\kappa-1}G'\,[2\coth\eta+(\kappa-1)h'/h+G''/G']$. With $G''/G'=u'/u+\lambda u$ and $h'/h=\lambda u-uu'/(1-u^2)$, which follows from $h=\lambda G\sqrt{1-u^2}$, division by $\lambda G\sinh^2\eta\,h^{\kappa-1}$ yields
\begin{equation}
u'=\frac{1-u^2}{1-\kappa u^2}\,N(\eta,u),\qquad N=\kappa(\lambda-\lH)-2u\coth\eta-\kappa\lambda u^2 .\label{eq:ueq}
\end{equation}
The sonic line is $\kappa u^2=1$, where the right-hand side diverges unless $N=0$. At the centre, regularity requires $u\simeq\kappa(\lambda-\lH)\,\eta/3$, so the regular branch has no free parameter.

The definitions~\eqref{eq:Adef}--\eqref{eq:Bdef} imply $B=A+2F'$ and $\lambda F=A+2sF'$, and these make several properties of the solution transparent. First, they give the identity
\begin{equation}
\lambda^2F^2=(1-s)H^2+sB^2 . \label{eq:identity}
\end{equation}
To see this, one has to substitute $H^2=A^2-4sF'^2$ and $B=A+2F'$ into the right-hand side of Eq.~\eqref{eq:identity}. The terms proportional to $sA^2$ and to $sF'^2$ cancel, and the remainder is $A^2+4sAF'+4s^2F'^2=(A+2sF')^2=\lambda^2F^2$. Using this identity, the denominator of Eq.~\eqref{eq:Fpp} factorises as
\begin{equation}
4s\left[(s-1)+\frac{(\kappa-1)sB^2}{H^2}\right]=4s\,\frac{\lambda^2F^2}{H^2}\left(\kappa v_{\rm rel}^2-1\right),\qquad
v_{\rm rel}^2\equiv\frac{sB^2}{\lambda^2F^2}. \label{eq:Dfact}
\end{equation}

Second, $v_{\rm rel}$ has a simple kinematic meaning. With $\xi=r/t=\sqrt s$, inserting the velocity~\eqref{eq:vF} and using Eq.~\eqref{eq:Bdef} together with Eq.~\eqref{eq:Adef} in the form $\lambda F=A+2sF'$ gives $\xi-v=\sqrt s\,B/A$ and $1-\xi v=\lambda F/A$, hence
\begin{equation}
\frac{\xi-v}{1-\xi v}=\frac{\sqrt s\,B}{\lambda F},\qquad\text{i.e.}\qquad |v_{\rm rel}|=\frac{|\xi-v|}{1-\xi v} . \label{eq:vrel}
\end{equation}
This is the relativistic velocity of the fluid measured by an observer moving along the similarity line $r=\xi t$, i.e.\ comoving with the self-similar pattern.

Third, $B$ itself measures how fast the fluid crosses this pattern. With $\partial_ts=-2s/t$, $\partial_rs=2\sqrt s/t$, $u^0=A/H$ and $u^r=v\,u^0=-2\sqrt s\,F'/H$, one finds
\begin{equation}
u^\mu\partial_\mu s=-\frac{2s}{tH}\,(A+2F')=-\frac{2s}{tH}\,B .
\end{equation}
For $B\equiv0$ every fluid element stays at fixed $s$, i.e.\ $v=r/t$ and $v_{\rm rel}\equiv0$. The condition $B\equiv0$ is the first-order equation $\lambda F+2(1-s)F'=0$, whose solution is $F=F(0)\,(1-s)^{\lambda/2}$. Eq.~\eqref{eq:FODE} then reduces to $\lambda(1-s)^{\lambda/2-1}\left[\kappa(\lambda-1)+3\right]=0$, which requires $\lambda=\lH$. Within the ansatz, Hubble flow is therefore the unique solution with $B\equiv0$; this also proves that $F_{\rm H}$ in Eq.~\eqref{eq:FH} solves the field equation.

The above also imply that the field equation is singular where $v_{\rm rel}=c_s=1/\sqrt\kappa$.  The physical origin of this singularity is the propagation of sound. In a radial flow, small disturbances travel along two families of characteristics, outgoing and incoming sound waves, with laboratory velocities
\begin{equation}
\frac{\dd r}{\dd t}=c_\pm\equiv\frac{v\pm c_s}{1\pm v\,c_s},
\end{equation}
i.e.\ the sound speed added to and subtracted from the local fluid velocity, relativistically. In a self-similar solution, a point of fixed shape of the profile, $\xi=r/t=\,$const, moves with velocity $\xi$. Measured from this moving point, a sound wave has velocity $(c_\pm-\xi)/(1-c_\pm\xi)$, which vanishes when $c_\pm=\xi$. Inserting the definition of $c_\pm$ shows that this happens exactly when the fluid moves with the sound speed relative to the pattern:
\begin{equation}
v\pm c_s=\xi\,(1\pm v\,c_s)
\;\;\Longleftrightarrow\;\;
\pm c_s\,(1-\xi v)=\xi-v
\;\;\Longleftrightarrow\;\;
\frac{\xi-v}{1-\xi v}=\pm c_s ,
\end{equation}
i.e.\ $|v_{\rm rel}|=c_s$ by Eq.~\eqref{eq:vrel}. Where $|v_{\rm rel}|<c_s$ (subsonic relative to the pattern), the two families cross each similarity line in opposite directions, so the profile at a given $\xi$ is influenced from both sides. Where $|v_{\rm rel}|>c_s$ (supersonic), both families cross in the same direction, and information flows across the pattern only one way. At $|v_{\rm rel}|=c_s$ one family of sound waves moves exactly with the pattern and does not cross the similarity line at all: the line $r=\xi t$ is then itself a characteristic. Data given along a characteristic do not determine the derivatives across it, which is why the coefficient of $F''$ in Eq.~\eqref{eq:Fpp} vanishes there, Eq.~\eqref{eq:Dfact}. A solution can then pass smoothly only if an extra regularity condition holds (the numerator vanishing as well, see below); otherwise the derivatives diverge, the profile steepens as in wave breaking, and a smooth self-similar solution cannot be continued. This is the relativistic analogue of the sonic (critical) point of Bondi accretion, the Parker wind, or the throat of a de Laval nozzle. Hubble flow has $v_{\rm rel}\equiv0$ and never becomes sonic.

A smooth solution can cross the sonic line only where the numerator of Eq.~\eqref{eq:Fpp} vanishes as well. One can show that these two conditions (vanishing numerator and denominator) fix a single point. Since the field equation is homogeneous in $F$, both conditions depend only on $s$ and on the ratio $F'/F$. The first condition can be solved for this ratio explicitly. Due to Eq.~\eqref{eq:Dfact}, the denominator vanishes when $\kappa sB^2=\lambda^2F^2$, i.e.\ when
\begin{equation}
B=\pm\frac{c_s\lambda F}{\xi},
\end{equation}
with $\xi=\sqrt s$. By Eq.~\eqref{eq:vrel}, the sign is that of $(\xi-v)/(1-\xi v)$ at the sonic point; here and below, the upper and lower signs belong together. Inserting $B$ from Eq.~\eqref{eq:Bdef} turns this into an explicit value of the logarithmic derivative of $F$,
\begin{equation}
\frac{F'}{F}=\frac{\lambda}{2(1-\xi^2)}\left(\pm\frac{c_s}{\xi}-1\right) .
\end{equation}
One can substitute the $F'$ from this into Eqs.~\eqref{eq:Adef} and~\eqref{eq:H2def}, to obtain
\begin{equation}
A=\lambda F\,\frac{1\mp c_s\xi}{1-\xi^2},\qquad H^2=\lambda^2F^2\,\frac{1-c_s^2}{1-\xi^2}.
\end{equation}
Substituting these expressions into the numerator of Eq.~\eqref{eq:Fpp}, multiplied by $H^2$, yields
\begin{equation}
\frac{\lambda^3F^3\,(1-c_s^2)}{c_s^2\,\xi\,(1-\xi^2)^2}\Big[\pm2c_s^3+\xi(1-\lambda)-\xi c_s^2(3-\lambda)\Big] .
\end{equation}
For $\lambda\neq0$ and $0<\xi<1$ the prefactor does not vanish, so the numerator vanishes only if the bracket does. Multiplying the bracket by $\kappa=c_s^{-2}$ gives the condition
\begin{equation}
\xi\left[(\kappa-1)\lambda-(\kappa-3)\right]=\pm2c_s .
\end{equation}
Since $\xi=r/t>0$, this condition can be satisfied only with the sign of $(\kappa-1)\lambda-(\kappa-3)$; the opposite sign would require $\xi<0$. Hence there is exactly one critical point. For $(\kappa-1)\lambda>\kappa-3$ the upper sign applies, i.e.\ the fluid moves slower than the pattern ($v<\xi$); for $(\kappa-1)\lambda<\kappa-3$ the lower sign applies, i.e.\ the fluid outruns the pattern ($v>\xi$). Squaring the condition gives its location in $s$:
\begin{equation}
s_c=\xi_c^2=\frac{4}{\kappa\left[(\kappa-1)\lambda-(\kappa-3)\right]^2} .\label{eq:sc}
\end{equation}

\subsection{Numerical behavior}
\begin{figure}[t]
\centering
\includegraphics[width=\textwidth]{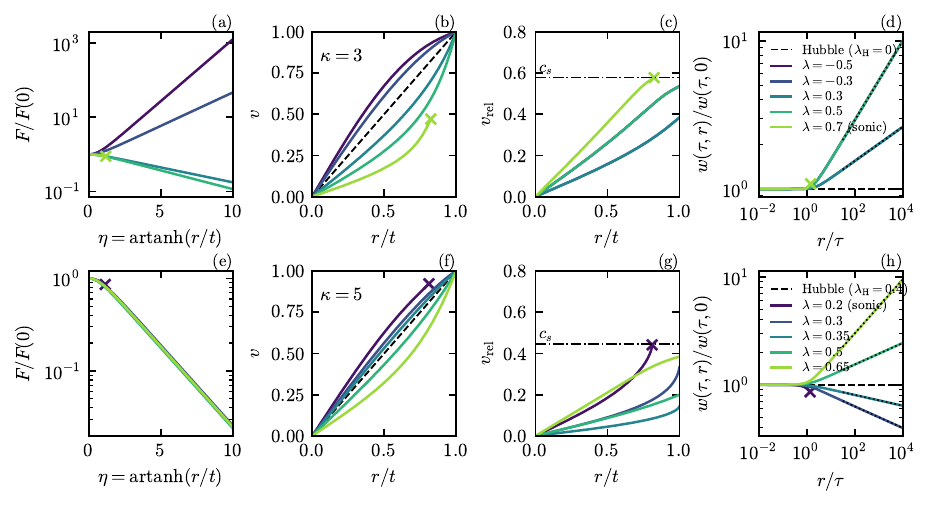}
\caption{Regular branch of \lPF{} for $\kappa=3$ (top row) and $\kappa=5$ (bottom row).
(a,e)~$F/F(0)$ versus the coordinate rapidity $\eta={\rm artanh}(r/t)$.
(b,f)~Radial velocity $v$ versus $r/t$; dashed: Hubble flow, $v=r/t$.
(c,g)~Fluid speed relative to the similarity lines, $v_{\rm rel}=\sqrt s\,|B|/|\lambda F|$; dash-dotted: $c_s=1/\sqrt\kappa$.
(d,h)~$w=p^{1/(\kappa+1)}$ on $\tau=$\,const, normalised at $r=0$; dotted lines show the asymptotic power law $(r/\tau)^\delta$ of Eq.~\eqref{eq:lightcone}.
Crosses mark where the regular branch reaches the sonic line $v_{\rm rel}=c_s$; the possible continuations beyond it are discussed in the text. For $\kappa=3$ the Hubble flow ($\Phi=\ln\tau$) lies outside the power ansatz and is shown only in (b) and (d).}
\label{fig:solI}
\end{figure}

\begin{figure}[t]
\centering
\includegraphics[width=0.75\columnwidth]{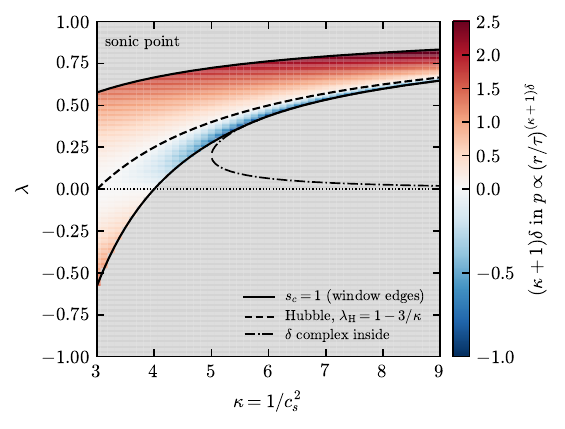}
\caption{Fate of the regular branch of \lPF{} in the $(\kappa,\lambda)$ plane. Grey: the regular branch reaches the sonic line before the light cone (cases (ii) and (iii) of the text). Solid lines: the analytic window edges, $s_c=1$ in Eq.~\eqref{eq:sc}; for $\kappa>3+2\sqrt2$ the lower edge is $\lambda_c$ of Eq.~\eqref{eq:lambdac}, the upper boundary of the complex-$\delta$ region (dash-dotted). Coloured: the branch reaches $s\to1$; the colour gives the exponent $(\kappa+1)\delta$ of the pressure on $\tau=$\,const at large $r/\tau$, Eq.~\eqref{eq:lightcone}. Dashed: Hubble flow, $\lH=1-3/\kappa$. Dash-dotted: boundary of the region where $\delta$ is complex (only for $\kappa>5$). Decaying tails (blue) occur only for $0<\lambda<\lH$, and in the scanned range they are never steeper than about $r^{-1}$.}
\label{fig:solIwindow}
\end{figure}

The regular branch is trivially subsonic in the origin, where $v_{\rm rel}=0$. Integrating Eq.~\eqref{eq:Fpp} outwards, we find three types of behavior, shown in Figs.~\ref{fig:solI} and~\ref{fig:solIwindow}. We discuss these as cases (i), (ii), and (iii) below.

\paragraph{(i) Subsonic window.} The regular branch stays subsonic relative to the pattern all the way to the light cone if and only if $\delta$ of Eq.~\eqref{eq:lightcone} is real and $|\delta/\lambda|<c_s$; there $v\to1$ and $v_{\rm rel}\to|\delta/\lambda|$. For $\kappa\le3+2\sqrt2\approx5.83$ this is equivalent to the critical point lying outside the light cone, $s_c>1$, i.e.\
\begin{equation}
\left|(\kappa-1)\lambda-(\kappa-3)\right|<2c_s .\label{eq:window}
\end{equation}
For $\kappa=3$ the window is $|\lambda|<1/\sqrt3$, and for $\kappa=5$ it is $0.276<\lambda<0.724$. For $\kappa>3+2\sqrt2$, Eq.~\eqref{eq:window} still gives the upper edge, but the lower edge is the boundary of the region where $\delta$ is complex,
\begin{equation}
\lambda_c=\frac12\left[\lH+\sqrt{\lH^2-4/\kappa^2}\right],\label{eq:lambdac}
\end{equation}
which lies below the lower edge $\lambda_-=(\kappa-3-2c_s)/(\kappa-1)$ of Eq.~\eqref{eq:window}, e.g.\ $0.533$ instead of $0.541$ for $\kappa=7$. The reason is as follows. At $\lambda_-$, the root of Eq.~\eqref{eq:lightcone} with $\delta/\lambda=-c_s$ has $1+\kappa\delta=1-\sqrt\kappa\,\lambda_-$, while the regular root has $1+\kappa\delta_+=+\sqrt{1+\kappa^2\lambda(\lambda-\lH)}>0$. The two coincide only if $\lambda_-<c_s$, i.e.\ $\kappa-3<(\kappa+1)c_s$, i.e.\ $(\sqrt\kappa+1)(\kappa-2\sqrt\kappa-1)<0$, i.e.\ $\kappa<3+2\sqrt2$. For larger $\kappa$, $\delta_+$ stays subsonic below $\lambda_-$, down to $\lambda_c$. Without this correction Eq.~\eqref{eq:window} would even exclude Hubble flow for $\kappa>9$, where $\lambda_->\lH$, although Hubble flow has $v_{\rm rel}\equiv0$. Numerically, for $3\le\kappa\le12$, the regular branch reaches the light cone exactly in this window, and the numerical edges agree with Eqs.~\eqref{eq:window} and~\eqref{eq:lambdac} to better than $2\times10^{-4}$ (solid lines in Fig.~\ref{fig:solIwindow}). Close to $\lambda_c$ this requires integrating far out (to $\eta\approx30$), since just below $\lambda_c$ the exponent $\delta$ has a small imaginary part and the branch becomes sonic only at large $\eta$. These solutions describe a flow emanating from a point and filling the interior of the light cone.

\paragraph{(ii) Transonic solutions.} The type of the critical point follows from Eq.~\eqref{eq:ueq}. Written as the autonomous system $\dd\eta/\dd\sigma=1-\kappa u^2$, $\dd u/\dd\sigma=(1-u^2)N$, it has a fixed point at $\kappa u_c^2=1$, $N=0$. There $\partial_\eta N=2u_c/\sinh^2\eta_c$, so the Jacobian has the determinant $4\kappa u_c^2(1-u_c^2)/\sinh^2\eta_c=4(1-c_s^2)/\sinh^2\eta_c>0$: the critical point is never a saddle, but a node or a focus. Its trace is $-2(1-c_s^2)(\coth\eta_c+\kappa\lambda u_c)$. With $\coth\eta_c=|X|/(2c_s)$ and $X\equiv(\kappa-1)\lambda-(\kappa-3)$ from Eq.~\eqref{eq:sc}, one finds $(\coth\eta_c+\kappa\lambda u_c)^2=\kappa[(\kappa+1)\lambda-(\kappa-3)]^2/4$ and $1/\sinh^2\eta_c=\kappa X^2/4-1$. The discriminant then factorises, and the critical point is a node if
\begin{equation}
(\lambda-1)\left[(3\kappa^2-6\kappa-1)\lambda-(3\kappa^2-14\kappa+7)\right]<0 ,\label{eq:node}
\end{equation}
and a focus otherwise. Above the window it is therefore a node for $\lambda<1$. Below the window it is a node only in a strip under $\lambda_-$, of width 0.42 for $\kappa=3$, 0.044 for $\kappa=4$ and 0.004 for $\kappa=5$, which closes at $\kappa=3+2\sqrt2$. Numerically, the regular branch runs into the critical point where it is a node, apart from a thin margin next to the node--focus boundary. Numerator and denominator of Eq.~\eqref{eq:Fpp} then vanish together, and $F''$ stays finite (e.g.\ $F''/F\approx-1.155$ at $s_c=100/147$ for $\kappa=3$, $\lambda=0.7$). For representative cases above the window ($\kappa=3$, $\lambda=0.6$--$0.9$; $\kappa=5$, $\lambda=0.75$--$0.9$) we continued the solution through $s_c$. The flow then becomes supersonic relative to the pattern and reaches $s\to1$ with $v<1$ and finite $w$ (e.g.\ $v\approx0.37$ for $\kappa=5$, $\lambda=0.8$). Such transonic solutions therefore require matter outside the light cone of the origin ($r>t$), i.e.\ initial data of infinite extent, as in the self-similar solutions with power-law initial data of Appendix~\ref{app:companion}. Moreover, the continuation through a nodal point is in general not unique. The continuation through the node, also below the window, is left for future work (Sec.~\ref{sec:summary}).

\paragraph{(iii) Fold singularity.} Where the critical point is a focus, i.e.\ below the node strip, and above the window for $\lambda>1$, the branch reaches $v_{\rm rel}=c_s$ with a non-vanishing numerator. There the denominator vanishes like $(s_\ast-s)^{1/2}$ and $F''\propto(s_\ast-s)^{-1/2}$ diverges: the profile folds over, similarly to wave breaking. The self-similar solution can then only be continued with a self-similar shock. This region contains the parameter range where $\delta$ given later in Eq.~\eqref{eq:lightcone} is complex.

In this paper we restrict ourselves to case~(i): solutions that stay subsonic relative to the similarity lines up to the light cone. They describe expansion from a point and are regular everywhere inside the light cone. The transonic solutions of case~(ii) and shock-connected solutions of case~(iii) are left for future work. 
%Finally, we note that for $\kappa=3$ the relative velocity and the $\tau=$\,const profile of $w$ are identical for $\pm\lambda$ (numerically to $10^{-9}$).

\subsection{Light-cone asymptotics}
For large $\eta$ we write $\Phi\simeq\tau^\lambda e^{\delta\eta}$. The leading-order behavior of Eq.~\eqref{eq:PhiPDE} gives
\begin{equation}
\kappa\lambda(\lambda-\lH)=\delta(2+\kappa\delta),\qquad
\delta=\frac{-1+\sqrt{1+\kappa^2\lambda(\lambda-\lH)}}{\kappa},\label{eq:lightcone}
\end{equation}
with $\delta_\pm=(-1\pm\sqrt D)/\kappa$ and $D=1+\kappa^2\lambda(\lambda-\lH)$. Equation~\eqref{eq:ueq} shows which root the regular branch selects. For large $\eta$, $\coth\eta\to1$, and $N=0$ is solved by $u=\delta_\pm/\lambda$, where $\partial N/\partial u=-2(1+\kappa\delta_\pm)=\mp2\sqrt D$. In the subsonic region, where $(1-u^2)/(1-\kappa u^2)>0$, deviations from $\delta_+/\lambda$ therefore decay, while deviations from $\delta_-/\lambda$ grow. Having no free parameter, the regular branch cannot be tuned onto the repelling root, and it approaches $\delta_+$, written as $\delta$ in Eq.~\eqref{eq:lightcone}. (For $\kappa<5$ the root $\delta_-$ would also be superluminal, $|\delta_-/\lambda|>1$, but not for $\kappa\ge5$: for $\kappa=5$, $D=(5\lambda-1)^2$ and $\delta_-=-\lambda$.)
%The numerical solutions confirm Eq.~\eqref{eq:lightcone} to $\sim10^{-6}$; for example $\kappa=5$, $\lambda=0.3$ gives $\delta=-0.1$.
On $\tau=$\,const the above implies
\begin{equation}
w\propto(r/\tau)^\delta,\qquad p\propto(r/\tau)^{(\kappa+1)\delta},\qquad \tanh\zeta\to-\delta/\lambda ,
\end{equation}
where $\zeta$ is the rapidity of the flow relative to Hubble flow (see Sec.~\ref{sec:solIaccel}). The consequences of these are the following.
\begin{itemize}
\item A decreasing pressure requires $\delta<0$, thus $0<\lambda<\lH=1-3/\kappa$, which is possible only for $\kappa>3$. For $\kappa=3$, every $\lambda\neq0$ gives an outward-increasing profile.
\item The decay is at most $p\propto r^{-(\kappa+1)/\kappa}$, and subsonicity restricts it further. With the isothermal or isentropic dressing, the particle number on $\tau=$\,const is infinite. Since $\dd^3\Sigma=\tau^3\sinh^2\eta\,\dd\eta\,\dd\Omega\simeq\tau r\,\dd r\,\dd\Omega$ for $r\gg\tau$, it diverges if $n$ falls more slowly than $r^{-2}$. Here $n\propto r^{(\kappa+1)\delta}$ or $r^{\kappa\delta}$, with $\kappa\delta\ge-1$. With the fast-tail dressing of Sec.~\ref{sec:obsSolI}, $n\propto r^{\kappa\delta-2\beta}$, and the number is finite if $2\beta>2+\kappa\delta$. This condition excludes a L\'evy-stable tail (Sec.~\ref{sec:obsSolI}).
\item For $\kappa>5$, $\delta$ becomes complex near the optimal $\lambda$. This region lies outside the window; for $\kappa>3+2\sqrt2$ its upper boundary is the lower window edge $\lambda_c$: there the regular branch ends at a fold singularity before reaching the light cone (case~(iii) above).
\end{itemize}
\lPF{} is therefore a regular, relativistic, accelerating family, but its pressure cannot carry an asymptotic L\'evy tail: it decays too slowly, and the outer flow is ultra-relativistic, so the slow-tail route is closed. Power-law sources can nevertheless arise through the fast-tail route, i.e.\ through the dressing on the freeze-out hypersurface. Since the pressure decays at most as $r^{-(\kappa+1)/\kappa}$, the bound~\eqref{eq:bound} leaves room for any steeper source tail. Section~\ref{sec:obsSolI} shows that the Hubble-type dressing gives L\'evy-stable tails for a range of parameters, although only with an infinite particle number.

\subsection{Acceleration}\label{sec:solIaccel}
We now quantify how strongly \lPF{} accelerates and how $\lambda$ controls it. The proper acceleration $a^\mu=u^\nu\partial_\nu u^\mu$ follows directly from Euler's equation~\eqref{eq:euler}: with $p=w^{\kappa+1}$ it reads $a^\mu=\Delta^{\mu\nu}\partial_\nu\ln w$. In a radial flow, $a^\mu$ is orthogonal to $u^\mu=\gamma(1,v)$ and therefore proportional to the outward unit vector $n^\mu=\gamma(v,1)$ of the $(t,r)$ plane, $a^\mu=a\,n^\mu$. The radial component of $\Delta^{\mu\nu}\partial_\nu$ is $-\partial_r-u^ru^\nu\partial_\nu=-\gamma^2(\partial_r+v\,\partial_t)$, and $n^r=\gamma$, so the magnitude of the acceleration, counted positive outwards, is
\begin{equation}
a=-\gamma\,(\partial_r+v\,\partial_t)\ln w .
\end{equation}
Inserting $\ln w=(\lambda-1)\ln t+\ln H(s)$, with $\partial_ts=-2s/t$ and $\partial_rs=2\sqrt s/t$, gives
\begin{equation}
t\,a=-\gamma\left[2\sqrt s\,(\ln H)'+v\left(\lambda-1-2s\,(\ln H)'\right)\right] .\label{eq:accel}
\end{equation}
The right-hand side depends on $s$ only: at fixed $r/t$ the acceleration decreases as $1/t$. Two limits show how $\lambda$ controls it.

Near the centre, $v\simeq c\,r/t$ with $c=\kappa(1-\lambda)/3$ (Sec.~\ref{sec:central}). A fluid element then moves as $r\propto t^c$, so its acceleration is $\dd^2r/\dd t^2=c(c-1)\,r/t^2$, which Eq.~\eqref{eq:accel} reproduces numerically. The core is therefore harmonic: the acceleration grows linearly with $r$, points outwards for $c>1$ ($\lambda<\lH$) and inwards for $c<1$, and vanishes for Hubble flow. The subsonic window bounds the possible values of $c$. For $\kappa\le3+2\sqrt2$ its edges are $\lambda_\pm=(\kappa-3\pm2c_s)/(\kappa-1)$, hence $1-\lambda_\pm=2(1\mp c_s)/(\kappa-1)$, and inserting these into $c=\kappa(1-\lambda)/3$ gives
\begin{equation}
\frac{2\kappa(1-c_s)}{3(\kappa-1)}<c<\frac{2\kappa(1+c_s)}{3(\kappa-1)} .\label{eq:cwindow}
\end{equation}
This means $0.42<c<1.58$ for $\kappa=3$ and $0.46<c<1.21$ for $\kappa=5$: a stiffer equation of state allows a stronger acceleration or deceleration of the core. For $\kappa>3+2\sqrt2$ the upper bound becomes $c<\kappa(1-\lambda_c)/3$, e.g.\ 1.089 instead of 1.072 for $\kappa=7$.

Far from the centre, the light-cone asymptotics determine the acceleration. In Milne coordinates $(\tau,\eta)$ we write the flow as $u^\mu=(\cosh\zeta,\sinh\zeta/\tau)$, where $\zeta$ is its rapidity relative to Hubble flow; the outward unit vector is then $n^\mu=(\sinh\zeta,\cosh\zeta/\tau)$. With $\Phi\simeq\tau^\lambda e^{\delta\eta}$, Eq.~\eqref{eq:potential} gives $w\cosh\zeta=\partial_\tau\Phi=\lambda\Phi/\tau$ and $-w\tau\sinh\zeta=\partial_\eta\Phi=\delta\Phi$, hence $\tanh\zeta=-\delta/\lambda$. Since $\ln w=(\lambda-1)\ln\tau+\delta\eta+{\rm const}$, the acceleration $a=-n^\mu\partial_\mu\ln w$ becomes
\begin{equation}
\tau a=-\left[(\lambda-1)\sinh\zeta+\delta\cosh\zeta\right]=-\cosh\zeta\left[\delta-(\lambda-1)\frac\delta\lambda\right]=-\frac{\delta}{\lambda}\cosh\zeta=-\frac{v_\infty}{\sqrt{1-v_\infty^2}} ,\label{eq:accelinf}
\end{equation}
where $v_\infty=\delta/\lambda$ is the limiting value of $v_{\rm rel}$, Eq.~\eqref{eq:vrel}, and $\cosh\zeta=(1-v_\infty^2)^{-1/2}$. On $\tau=$\,const the acceleration in the outer region therefore tends to a constant, fixed by $\lambda$. Within the window $|v_\infty|<c_s$, so $|\tau a|<c_s/\sqrt{1-c_s^2}=1/\sqrt{\kappa-1}$.

\begin{figure*}[t]
\centering
\includegraphics[width=\textwidth]{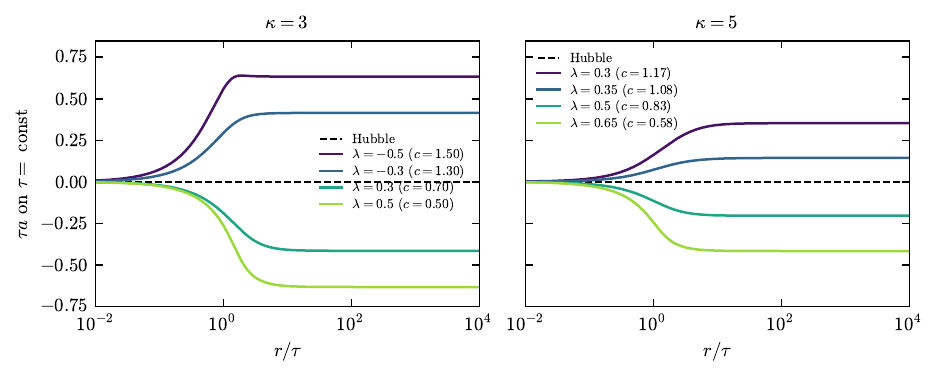}
\caption{Proper acceleration of \lPF{} on $\tau=$\,const, in units of $1/\tau$, for $\kappa=3$ (left) and $\kappa=5$ (right). Positive values point outwards; dashed: Hubble flow ($a=0$). Close to the centre, $\tau a\simeq c(c-1)\,r/\tau$ with $c=\kappa(1-\lambda)/3$; far from it, $\tau a$ tends to the constant of Eq.~\eqref{eq:accelinf}.}
\label{fig:solIaccel}
\end{figure*}

Figure~\ref{fig:solIaccel} shows both regimes; the numerical solutions reproduce the core and outer limits to better than $10^{-3}$. In contrast to the slow-tail criterion~\eqref{eq:criterion}, the acceleration of \lPF{} does not fall off as $1/r$. In the outer region the flow stays ultra-relativistic, with a constant rapidity offset $\zeta$ relative to Hubble flow. The \lPF{} solutions therefore belong to the fast-tail class of Sec.~\ref{sec:mismatch}, which is why its sources behave like those of Hubble flow (Sec.~\ref{sec:obsSolI}).

%=====================================================================
\section{Gubser flow with comoving dressing}
\label{sec:solII}
%=====================================================================

\subsection{Definitions}
We use Milne coordinates $\tau=\sqrt{t^2-z^2}$, $\eta_s={\rm artanh}(z/t)$, the transverse radius $r$ and azimuth $\varphi$, with $\dd s^2=\dd\tau^2-\dd r^2-r^2\dd\varphi^2-\tau^2\dd\eta_s^2$. The EoS is conformal, $\varepsilon=3p$ and $p=nT$. The parameters are the transverse scale $q$, a normalisation $\hat\varepsilon_0$, and the freeze-out values $\tau_f$ and $T_f$. With
\begin{equation}
\mathcal{D}\equiv1+2q^2(\tau^2+r^2)+q^4(\tau^2-r^2)^2,
\end{equation}
Gubser's solution~\cite{Gubser:2010ze} reads
\begin{equation}
u^\tau=\cosh\chi,\quad u^r=\sinh\chi,\quad u^\varphi=u^\eta=0,\qquad
\tanh\chi=v_\perp=\frac{2q^2\tau r}{1+q^2\tau^2+q^2r^2},
\end{equation}
\begin{equation}
p=\frac{\hat\varepsilon_0}{3\tau^4\cosh^{8/3}\!\rho}=\frac{\hat\varepsilon_0(2q)^{8/3}}{3\tau^{4/3}\mathcal{D}^{4/3}} ,
\end{equation}
with the de Sitter coordinates
\begin{equation}
\sinh\rho=-\frac{1-q^2\tau^2+q^2r^2}{2q\tau},\quad \cosh\rho=\frac{\sqrt{\mathcal{D}}}{2q\tau},\quad
\cot\theta=\frac{1+q^2\tau^2-q^2r^2}{2qr},\ \theta\in[0,\pi] ,
\end{equation}
and their inverse
\begin{equation}
q\tau=\frac{1}{\cosh\rho\cos\theta-\sinh\rho},\qquad qr=\frac{\sin\theta\cosh\rho}{\cosh\rho\cos\theta-\sinh\rho} .
\end{equation}
The angle $\theta$ (together with $\varphi$ and $\eta_s$) is constant along streamlines, $u^\mu\partial_\mu\theta=0$.

\subsection{Dressing}
By the dressing theorem, any $\nu(\theta)>0$ gives an exact solution with conserved particle number:
\begin{equation}
n=p^{3/4}\,\nu(\theta),\qquad T=\frac{p^{1/4}}{\nu(\theta)} .
\end{equation}
The isentropic dressing, $\nu=$\,const, gives Gubser's original solution, with $T\propto p^{1/4}$. The \emph{isothermal dressing}, $T(\tau_f,r)=T_f$, is
\begin{equation}
T(\tau,r)=T_f\,\frac{\tau_f}{\tau}\left[\frac{\cosh\rho_f(\theta)}{\cosh\rho(\tau,r)}\right]^{2/3},\qquad n=\frac pT ,\label{eq:Tiso}
\end{equation}
where $\rho_f(\theta)$ is evaluated on the freeze-out surface along the same streamline:
\begin{align}
q\,r_f(\theta)&=-\cot\theta+\sqrt{\cot^2\theta+1+q^2\tau_f^2},\\
\cosh\rho_f&=\frac{\sqrt{\mathcal{D}(\tau_f,r_f)}}{2q\tau_f}
\quad\Longleftrightarrow\quad
\cosh\rho_f\cos\theta-\sinh\rho_f=\frac{1}{q\tau_f}.
\end{align}

\subsection{Properties}

Let us now discuss a few properties of the Gubser solution. We first observe that on $\tau=\tau_f$, $n=p/T_f\propto \mathcal{D}(\tau_f,r)^{-4/3}$ holds. This also indicates that for $q\tau_f>1$ the pressure peaks on a ring at $r_\ast=\sqrt{\tau_f^2-q^{-2}}$. To give an example, for $q^{-1}=4.3$\,fm and $\tau_f=8$\,fm, $r_\ast\approx6.7$\,fm. Furthermore, at large $r$ on $\tau=$\,const one has $\mathcal{D}\simeq q^4r^4$, which simply yields $p\propto r^{-16/3}$ and $v_\perp\to2\tau/r$. Since the Gubser flow is boost invariant, the relevant source lives in the two-dimensional transverse plane (Sec.~\ref{sec:obs}), where a L\'evy-stable tail would decrease as $r^{-(2+\alpha)}$ (Sec.~\ref{sec:exp}). Matching the pressure tail exponent to this form gives $2+\alpha_\infty=16/3$, thus the asymptotic exponent is $\alpha_\infty=10/3$, which does not correspond to a Lévy-stable distribution. On the other hand, the numerically calculated observables may still appear to be Lévy-like over a finite range, as we will discuss subsequently.

Before that, let us also discuss the acceleration of this solution. The proper acceleration decreases as $1/r$, as required by the slow-tail route, but its coefficient differs from the quasi-static value of Eq.~\eqref{eq:criterion}. In Milne coordinates at $\eta_s=0$, Eq.~\eqref{eq:aslow} with $\kappa=3$ reads $4a=-\cosh\chi\,(\partial_r\ln p+v_\perp\partial_\tau\ln p)$, and $\cosh\chi\to1$. The pressure gradient gives $-\partial_r\ln p\simeq16/(3r)$. For the second term, $p\propto\tau^{-4/3}\mathcal{D}^{-4/3}$ and $\partial_\tau\ln \mathcal{D}=-4\tau/r^2+\mathcal{O}(r^{-4})$, so $-\partial_\tau\ln p\simeq4/(3\tau)$; multiplied by $v_\perp\simeq2\tau/r$, this gives $8/(3r)$. Hence
\begin{equation}
4a\simeq\frac{16}{3r}+\frac{8}{3r}=\frac{8}{r},\qquad\text{i.e.}\qquad a\simeq\frac{2}{r} ,
\end{equation}
instead of the quasi-static value $4/(3r)$.
%The same result follows directly from the kinematics, $a\simeq\partial_\tau v_\perp=2/r$, and we verified numerically that $r\,a\to2$ on $\tau=\tau_f$.

The maximal transverse velocity for $\tau=$\,const is $v_{\max}=q\tau/\sqrt{1+q^2\tau^2}$, this is reached at $r=\sqrt{\tau^2+q^{-2}}$. To give a numerical example, for $q^{-1}=4.3$\,fm and $\tau_f=8$\,fm, the maximal transverse velocity is $v_{\max}=0.88$, reached at $r=9.1$\,fm.

\begin{figure}[t]
\centering
\includegraphics[width=0.5\columnwidth]{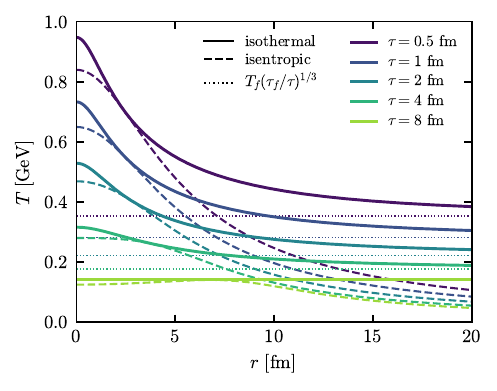}
\caption{Temperature of the dressed Gubser flow at early times, $q^{-1}=4.3$\,fm, $\tau_f=8$\,fm, $T_f=140$\,MeV. Solid: isothermal dressing, Eq.~\eqref{eq:Tiso}. Dashed: isentropic dressing, $\nu=$\,const, normalised to $T=T_f$ at the pressure maximum on $\tau_f$. Dotted: large-$r$ limit of the isothermal dressing, $T_f(\tau_f/\tau)^{1/3}$.}
\label{fig:Tearly}
\end{figure}

The isothermal dressing fixes the entropy per particle on each streamline by its value at freeze-out, which determines the temperature at earlier times (Fig.~\ref{fig:Tearly}). Since $\nu$ is constant along streamlines and $p$ decreases along each of them, every fluid element cools monotonically and reaches $T_f$ exactly on $\tau_f$. In the origin, Eq.~\eqref{eq:Tiso} gives
\begin{equation}
T(\tau,0)=T_f\left(\frac{\tau_f}{\tau}\right)^{1/3}\left(\frac{1+q^2\tau_f^2}{1+q^2\tau^2}\right)^{2/3} .
\end{equation}
Far from the centre the transverse flow vanishes due to $v_\perp\simeq2\tau/r$,thus the streamlines stay at nearly constant $r$, and each fluid element cools only through the longitudinal expansion, as in Bjorken flow: $T\to T_f(\tau_f/\tau)^{1/3}$ for $r\to\infty$. The isothermal dressing therefore implies a hot, dilute periphery at early times. At $\tau=1$\,fm, for example, the temperature drops from 0.73\,GeV in the centre only to 0.29\,GeV at $r=40$\,fm, whereas with the isentropic dressing (normalised to $T=T_f$ at the pressure maximum on $\tau_f$) it drops to 0.03\,GeV. Within the perfect-fluid model this is harmless: the dynamics depends only on $p$, and the energy density is the same for every dressing. It would gain more relevance and would be thus more problematic in extensions of this model in which the temperature enters the dynamics, e.g.\ through a temperature-dependent equation of state or viscosity.

%=====================================================================
\section{Observables}
\label{sec:obs}
%=====================================================================

In this section we compute the single-particle spectra and the femtoscopic observables of the solutions discussed above. The three solutions enter the section at different depths, for the following reasons.
\begin{itemize}
\item The \emph{Hubble-type L\'evy solution} is not new. Its spectra, sources and L\'evy fits were computed in Ref.~\cite{Szentkuti:2026buv} and are summarised in Sec.~\ref{sec:hubble}. Here it serves only as a reference: it is the special case $\lambda=\lH$ of \lPF{} with the fast-tail dressing (Subsec.~\ref{sec:obsSolI}).
\item The \emph{dressed Gubser flow} is the solution of this paper that realises the slow-tail route with a regular centre and a finite source size. Its boost invariance and azimuthal symmetry allow the $\eta_s$ and $\varphi$ integrals to be done analytically, so all observables reduce to one-dimensional radial integrals or closed-form expressions. We therefore compute the full set of observables for it: spectra (Subsec.~\ref{sec:gubserspectra}), the source function and its tail (Subsec.~\ref{sec:gubsertail}), Gaussian HBT radii (Subsec.~\ref{sec:gubserradii}), and L\'evy fits (Subsec.~\ref{sec:levyfits}).
\item \lPF{} has no power-law pressure tail of its own (Sec.~\ref{sec:solI}). With the isothermal or isentropic dressing it has no geometric scale (such as $R_{\rm geom}$ relevant for femtoscopy) besides $\tau_f$, and its sources are far from the measured ones. With the fast-tail dressing it generalises the Hubble-type mechanism to an accelerating core. For \lPF{} we therefore concentrate on the source function, i.e.\ on its tail and on L\'evy fits (Subsec.~\ref{sec:obsSolI}).
\end{itemize}
Subsection~\ref{sec:tailsummary} finally compares the source tails of the three solutions.

\subsection{Dressed Gubser flow: single-particle spectra}\label{sec:gubserspectra}
For the dressed Gubser flow with freeze-out at $\tau=\tau_f$ we have $K^\mu\dd\Sigma_\mu=m_T\cosh(y-\eta_s)\,\tau_fr\,\dd r\,\dd\varphi\,\dd\eta_s$ and
\begin{equation}
K\cdot u=m_T\cosh(y-\eta_s)\cosh\chi-K_T\cos(\varphi-\varphi_K)\sinh\chi .
\end{equation}
The $\eta_s$ and $\varphi$ integrals can be done analytically, giving a blast-wave form~\cite{Schnedermann:1993ws} with the Gubser profile:
\begin{equation}
E\frac{\dd N}{\dd^3K}=\frac{g\,\tau_f\,m_T}{2\pi^2\,\Teq(T_f)}\int_0^\infty r\,\dd r\;n(\tau_f,r)\;
K_1\!\left(\frac{m_T\cosh\chi}{T_f}\right)I_0\!\left(\frac{K_T\sinh\chi}{T_f}\right) .\label{eq:spectrum}
\end{equation}
For the benchmark parameters, $q^{-1}=4.3$\,fm~\cite{Gubser:2010ze}, $\tau_f=8$\,fm and $T_f=140$\,MeV, the transverse flow at freeze-out is very strong: $v_{\max}=0.88$, and the density-weighted mean velocity on $\tau_f$ is $\langle v_\perp\rangle=0.79$. The pion spectrum is then nearly exponential, but with an effective temperature $T_{\rm eff}\approx0.46$\,GeV, considerably above the measured value of about 0.2\,GeV~\cite{PHENIX:2003iij}. Here and below, $T_{\rm eff}$ is defined as the negative inverse slope of a linear fit of $\ln E\,\dd N/\dd^3K$ versus $m_T$ over $m_T-m=0.1$--$1.0$\,GeV. For kaons and protons the flow even reverses the slope at low $m_T$: the spectrum rises with $m_T$ up to $m_T-m\approx0.22$\,GeV for kaons and 0.6\,GeV for protons, so that no effective temperature can be defined. The non-relativistic estimate $T_{\rm eff}\simeq T_f+m\langle v_\perp^2\rangle$ does not apply at such velocities. At high $m_T$ the slope approaches the blue-shifted value $T_f\sqrt{(1+v_{\max})/(1-v_{\max})}$, here 0.56\,GeV.

The strength of the flow on $\tau_f$ is set by $q\tau_f$ alone. Table~\ref{tab:gubserscan} shows a scan in $q$ at fixed $\tau_f=8$\,fm. For $q^{-1}=10$--$12$\,fm, i.e.\ $q\tau_f=0.67$--$0.8$, the transverse flow is moderate ($v_{\max}=0.55$--$0.62$), the pion effective temperature is 0.22--0.25\,GeV, close to the measured value, and the effective temperatures increase with the particle mass, as expected from radial flow and as observed~\cite{PHENIX:2003iij}. For $q\tau_f<1$ the pressure on $\tau_f$ peaks at the centre instead of on a ring. Whether the L\'evy-like shape of the source survives for this weaker flow is examined in Subsec.~\ref{sec:levyfits}.

For comparison, the Hubble-type L\'evy solution gives pion effective temperatures of about 0.3--0.5\,GeV in the parameter range of Ref.~\cite{Szentkuti:2026buv} (Sec.~\ref{sec:hubble}), similar to the benchmark Gubser flow. For \lPF{} we do not compute spectra, for the reasons given at the beginning of this section.

\begin{table}[t]
\caption{Dressed Gubser flow with the isothermal dressing at $\tau_f=8$\,fm and $T_f=140$\,MeV for several values of $q^{-1}$: maximal and mean (density-weighted) transverse velocity on $\tau_f$, and effective temperatures (GeV) of pions, kaons and protons from linear fits of $\ln E\,\dd N/\dd^3K$ over $m_T-m=0.1$--$1.0$\,GeV. A dash means that the spectrum still rises with $m_T$ above $m_T-m=0.1$\,GeV, i.e.\ inside the fit range (see text). $q^{-1}=4.3$\,fm corresponds to the benchmark parameters.}
\label{tab:gubserscan}
\begin{ruledtabular}
\begin{tabular}{cccccc}
$q^{-1}$ (fm) & $v_{\max}$ & $\langle v_\perp\rangle$ & $T_{\rm eff}^\pi$ & $T_{\rm eff}^K$ & $T_{\rm eff}^p$\\
\hline
4.3 & 0.88 & 0.79 & 0.46 & -- & --\\
6 & 0.80 & 0.70 & 0.35 & 0.53 & --\\
8 & 0.71 & 0.60 & 0.28 & 0.37 & 0.55\\
10 & 0.62 & 0.52 & 0.25 & 0.29 & 0.38\\
12 & 0.55 & 0.46 & 0.22 & 0.26 & 0.31\\
\end{tabular}
\end{ruledtabular}
\end{table}

\subsection{Dressed Gubser flow: source function and its tail}\label{sec:gubsertail}
For the isothermal dressing, integrating Eq.~\eqref{eq:CF} over $\eta_s$ at $y=0$ gives the transverse source
\begin{equation}
S_\perp(r,\varphi;K)\propto n(\tau_f,r)\,K_1\!\left(\frac{m_T\cosh\chi}{T_f}\right)\exp\!\left[\frac{K_T\sinh\chi\cos(\varphi-\varphi_K)}{T_f}\right] .\label{eq:Sperp}
\end{equation}
For large $r$, $\chi\to0$, so $S_\perp\propto r^{-16/3}K_1(m_T/T_f)$ for every $K$: a power-law tail. In two transverse dimensions this corresponds to an asymptotic index $\alpha_\infty=10/3$, outside the stable range. The source is therefore L\'evy-like only over a finite range, which is the range probed by the femtoscopic (quantum-statistical) correlation measurements.

Could another dressing of the same flow give a L\'evy-stable tail? To answer this we need the source for a general dressing. The $\eta_s$ integral is $\int\dd\eta_s\cosh\eta_s\,e^{-a\cosh\eta_s}=2K_1(a)$ with $a=m_T\cosh\chi/T$. For a general dressing, $n(\tau_f,r)$ in Eq.~\eqref{eq:Sperp} is therefore replaced by the fugacity, and $T_f$ by $T(\tau_f,r)$:
\begin{equation}
S_\perp\propto\frac{n}{\Teq(T)}\,K_1\!\left(\frac{m_T\cosh\chi}{T}\right)\exp\!\left[\frac{K_T\sinh\chi\cos(\varphi-\varphi_K)}{T}\right] .\label{eq:Sperpgen}
\end{equation}
For dressings with a freeze-out temperature bounded away from zero, $T\geq T_{\min}>0$, the inequality $xK_1(x)\leq1$, i.e.\ $K_1(m_T\cosh\chi/T)\leq T/(m_T\cosh\chi)\leq T/m_T$, gives
\begin{equation}
S_\perp\lesssim\frac{T\,p}{p_{\rm eq}(T)}
=\frac{p}{\Teq(T)} \leq\frac{p}{\Teq(T_{\min})} \propto r^{-16/3}.
\label{eq:gubserbound}
\end{equation}
Here we used the monotonic increase of $\Teq(T)$ with temperature. We also dropped the angular exponential of Eq.~\eqref{eq:Sperpgen}. It tends to unity in the outer region, since there $\sinh\chi\simeq v_\perp\simeq2\tau_f/r$ while $T\geq T_{\min}>0$, and therefore does not affect the asymptotic bound. The isothermal dressing saturates this bound, so no dressing with a positive lower temperature bound can produce a L\'evy-stable asymptotic tail.

The case in which $T$ tends to zero requires a separate argument. For a massive Boltzmann gas, the low-temperature asymptotics are $p_{\rm eq}(T)\propto T^{5/2}e^{-m/T}$, so that the fugacity is $p/p_{\rm eq}(T)\propto p\,T^{-5/2}e^{m/T}$, and $K_1(A)\propto A^{-1/2}e^{-A}$, so that $K_1(m_T\cosh\chi/T)\propto T^{1/2}e^{-m_T\cosh\chi/T}$. Here the angular exponential of Eq.~\eqref{eq:Sperpgen} cannot be dropped, because $1/T$ grows in the tail. Combining the three exponentials, we obtain
\begin{equation}
S_\perp\propto p\,T^{-2}\exp\left(-\frac{\Delta}{T}\right),\qquad
\Delta=m_T\cosh\chi-K_T\sinh\chi\cos(\varphi-\varphi_K)-m .\label{eq:gubserlowT}
\end{equation}
Compared with the three-dimensional source of Sec.~\ref{sec:bound}, the $\eta_s$ integral contributes the factor $T^{1/2}$ of $K_1$, so the power $T^{-5/2}$ found there becomes $T^{-2}$ here. Since $\chi\to0$ in the outer region, $\Delta\to m_T-m$. As a function of $T$, the factor $T^{-2}e^{-\Delta/T}$ is maximal where $\dd(-2\ln T-\Delta/T)/\dd T=-2/T+\Delta/T^2=0$, i.e.\ at $T=\Delta/2$, so $S_\perp\lesssim4e^{-2}p/\Delta^2$. For $m_T>m$ the source is therefore again bounded by the pressure, with a constant that diverges as $K_T\to0$, as in Sec.~\ref{sec:bound}. In particular, for the isentropic Gubser dressing, $T\propto p^{1/4}\propto r^{-4/3}$. Since $1/T\propto r^{4/3}$, $\ln S_\perp\simeq-c\,r^{4/3}$ with $c>0$ for $m_T>m$: the tail falls faster than exponentially (a compressed exponential). The angular term in $\Delta/T$ grows only as $K_T\sinh\chi/T\propto r^{-1+4/3}=r^{1/3}$ and does not change this. Only at exactly $K_T=0$ is there no such suppression: then $\Delta=m(\cosh\chi-1)\simeq mv_\perp^2/2\propto r^{-2}$, so $\Delta/T\propto r^{-2/3}\to0$, and $S_\perp\propto p\,T^{-2}\propto r^{-16/3+8/3}=r^{-8/3}$, which would correspond to $\alpha_\infty=2/3$. Thus, for any fixed nonzero $K_T$, no regular dressing of Gubser flow can produce an asymptotically L\'evy-stable tail.

\subsection{Dressed Gubser flow: Gaussian HBT radii}\label{sec:gubserradii}
For the dressed Gubser flow the saddle-point radii of Eq.~\eqref{eq:homog} can be written in closed form. (For \lPF{}, which is known only numerically, we characterise the sources directly by L\'evy fits in Subsec.~\ref{sec:obsSolI}.) For the source~\eqref{eq:Sperp} we use the leading-exponential approximation to $K_1$ and define
\begin{equation}
\Psi(r)=\ln n(\tau_f,r)-\frac{m_T\cosh\chi(r)-K_T\sinh\chi(r)}{T_f} .
\end{equation}
The emission maximum lies on the $K_T$ axis at $r_s$, with $\Psi'(r_s)=0$. Applying Eq.~\eqref{eq:homog} gives
\begin{equation}
R_{\rm side}^{-2}=\frac{K_T\sinh\chi(r_s)}{T_f\,r_s^2},\qquad
R_{\rm out}^{-2}=-\Psi''(r_s),\qquad
R_{\rm long}^{2}=\frac{\tau_f^2\,T_f}{m_T\cosh\chi(r_s)} .\label{eq:radii}
\end{equation}
The side formula is exact at the Gaussian level, because the saddle condition cancels the radial-curvature terms. For a linear flow, $\sinh\chi=Hr$, and a Gaussian profile, the saddle condition reads $K_TH/T_f=r_s/R_{\rm geom}^2+m_TH^2r_s/(T_f\cosh\chi)$, so $R_{\rm side}^{-2}=R_{\rm geom}^{-2}+m_TH^2/(T_f\cosh\chi(r_s))$, which is Eq.~\eqref{eq:CsL} to leading order in $Hr_s$; and $R_{\rm long}$ reproduces the Makhlin--Sinyukov form~\cite{Makhlin:1987gm} with transverse flow. Table~\ref{tab:gauss} lists the values for $q^{-1}=4.3$\,fm, $\tau_f=8$\,fm and $T_f=140$\,MeV. The emission maximum sits near the pressure ring, where $v_\perp\approx0.83$--$0.87$.

In this approximation we utilized only the exponential dependence $K_1(A)\propto e^{-A}$, neglecting the algebraic prefactor $A^{-1/2}$ and higher-order corrections in $1/A$. The resulting Gaussian homogeneity lengths should therefore be understood as leading-exponential saddle-point estimates; using the full Bessel function would shift the saddle position and modify the radial curvature, particularly for $R_{\rm out}$. Moreover, $R_{\rm out}$ in Eq.~\eqref{eq:radii} is the spatial homogeneity length. The measured out radius also contains the emission duration, which on $\tau=\tau_f$ arises because the emission time $t=\tau_f\cosh\eta_s$ varies with $\eta_s$ (see the caption of Table~\ref{tab:gauss}).

\begin{table}[t]
\caption{Gaussian homogeneity lengths, Eq.~\eqref{eq:radii}, for the isothermally dressed Gubser flow ($q^{-1}=4.3$\,fm, $\tau_f=8$\,fm, $T_f=140$\,MeV, pions), in the leading-exponential approximation to $K_1$. At $m_T=0.2$\,GeV the exact $K_1$ gives $r_s=5.2$\,fm, $R_{\rm side}=4.8$\,fm, $R_{\rm out}=4.2$\,fm and $R_{\rm long}=5.4$\,fm; at $m_T=0.8$\,GeV the differences are below 7\%. $R_{\rm out}$ is the spatial out homogeneity length. The emission duration, evaluated from the second moments of the full source as $\beta_T^2\,{\rm Var}(t)-2\beta_T\,{\rm Cov}(x_{\rm out},t)$ with $\beta_T=K_T/m_T$ and $t=\tau_f\cosh\eta_s$, adds 6.5, 1.4 and 0.45\,fm$^2$ to $R_{\rm out}^2$ at $m_T=0.2$, 0.5 and 0.8\,GeV. Lengths are in fm.}
\label{tab:gauss}
\begin{ruledtabular}
\begin{tabular}{ccccccc}
$m_T$ (GeV) & $r_s$ & $v_\perp(r_s)$ & $R_{\rm side}$ & $R_{\rm out}$ & $R_{\rm long}$ & $R_{\rm side}^{-2}$ (fm$^{-2}$)\\
\hline
0.2 & 6.32 & 0.826 & 5.16 & 3.29 & 5.03 & 0.038\\
0.3 & 6.91 & 0.849 & 3.96 & 2.77 & 3.97 & 0.064\\
0.4 & 7.16 & 0.857 & 3.40 & 2.54 & 3.40 & 0.087\\
0.5 & 7.33 & 0.861 & 3.04 & 2.40 & 3.02 & 0.108\\
0.6 & 7.45 & 0.864 & 2.79 & 2.29 & 2.74 & 0.129\\
0.8 & 7.63 & 0.868 & 2.44 & 2.14 & 2.36 & 0.169\\
\end{tabular}
\end{ruledtabular}
\end{table}

\subsection{Dressed Gubser flow: L\'evy fits}\label{sec:levyfits}
\begin{figure}[t]
\centering
\includegraphics[width=0.7\columnwidth]{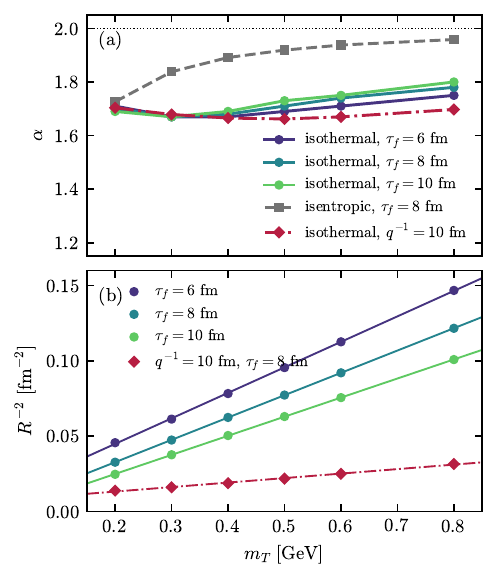}
\caption{L\'evy fits for the dressed Gubser flow ($T_f=140$\,MeV, pions), side direction. (a)~L\'evy exponent $\alpha$ versus $m_T$ for $q^{-1}=4.3$\,fm with the isothermal dressing at $\tau_f=6$, 8 and 10\,fm and with the isentropic dressing at $\tau_f=8$\,fm, and for $q^{-1}=10$\,fm with the isothermal dressing at $\tau_f=8$\,fm (diamonds); values of Table~\ref{tab:levy}. (b)~$R^{-2}$ versus $m_T$ for the isothermal dressing; points are the fitted values, lines the linear fits of Table~\ref{tab:levy}.}
\label{fig:levy}
\end{figure}

Beyond the Gaussian approximation we compute $C(Q)-1=|\tilde S_\perp(Q)|^2/|\tilde S_\perp(0)|^2$ for $Q$ in the side direction and fit $e^{-|QR|^\alpha}$ over $0.1<C-1<0.9$ (Appendix~\ref{app:numerics}). The same procedure is applied to \lPF{} in Subsec.~\ref{sec:obsSolI}. The results are given in Table~\ref{tab:levy}, Fig.~\ref{fig:levy}, and the following list.
\begin{itemize}
\item With the isothermal dressing, $\alpha\approx1.67$--$1.80$. It increases with $m_T$ above $m_T\approx0.3$\,GeV, qualitatively as in the recent data.
\item $R^{-2}$ is linear in $m_T$ to high accuracy. The slope and intercept depend on $\tau_f$, as shown in Table~\ref{tab:levy}.
\item With the isentropic dressing, $\alpha$ approaches 2 (1.89--1.96) for $m_T\gtrsim0.4$\,GeV, as expected from the compressed-exponential tail.
\item The L\'evy-like shape is not a consequence of the strong flow of the benchmark parameters. For the $q$ scan of Table~\ref{tab:gubserscan} ($q^{-1}=4.3$--$12$\,fm at $\tau_f=8$\,fm), the pion $\alpha$ at $m_T=0.2$, 0.5 and 0.8\,GeV stays between 1.66 and 1.78 for every $q$. The slope $a$ of a linear fit $R^{-2}=a\,(m_T+b)$ to these three points decreases with the flow, from 0.148 ($q^{-1}=4.3$\,fm) to 0.088, 0.050, 0.030 and 0.018\,fm$^{-2}$\,GeV$^{-1}$ ($q^{-1}=6$, 8, 10, 12\,fm), as expected from Eq.~\eqref{eq:CsL}, in which the slope is set by the flow gradient.
\item For $q^{-1}=10$\,fm, which gives nearly realistic spectra (Table~\ref{tab:gubserscan}), $\alpha=1.66$--$1.70$ for $m_T=0.2$--$0.8$\,GeV, with a minimum near $m_T=0.5$\,GeV and an increase above it (1.73 at $m_T=1$\,GeV). The radii are larger, $R=8.5$--$5.6$\,fm, and $R^{-2}=0.030\,(m_T+0.25\,{\rm GeV})$\,fm$^{-2}$ describes them to within 5\%. The L\'evy-like shape and the linear $R^{-2}(m_T)$ thus persist for realistic flow, while the increase of $\alpha$ with $m_T$ becomes weaker.
\item Pions and kaons with the same $m_T$ do not give identical sources. Exact $m_T$ scaling would require Eq.~\eqref{eq:Sperp} to depend on $K$ only through $m_T$, but it also depends on $K_T=\sqrt{m_T^2-m^2}$, which at a given $m_T$ is smaller for kaons. For the benchmark parameters at $m_T=0.6$\,GeV, the kaon source has $\alpha=1.40$ and $R=3.8$\,fm, compared with $\alpha=1.74$ and $R=3.3$\,fm for pions. The difference decreases as $m_T$ grows beyond the kaon mass; at $m_T=1.5$\,GeV it is $0.04$ in $\alpha$ and 3\% in $R$. For $q^{-1}=10$\,fm it is smaller: at $m_T=0.6$--$1.0$\,GeV the kaon $\alpha$ is lower by 0.05--0.08 and $R$ is larger by 5--14\%. Approximate $m_T$ scaling across species, as seen in the data~\cite{Kovacs:2023qax,Mukherjee:2023hrz}, is thus reproduced only for moderate flow or at large $m_T$.
\end{itemize}

\begin{table}[t]
\caption{L\'evy fits (side direction) for the dressed Gubser flow, $T_f=140$\,MeV, pions, at $m_T=0.2,0.3,0.4,0.5,0.6,0.8$\,GeV, with $q^{-1}=4.3$\,fm unless stated otherwise. The last column is a linear fit $R^{-2}=a\,(m_T+b)$.}
\label{tab:levy}
\begin{ruledtabular}
\begin{tabular}{lll}
Case & $\alpha(m_T)$ & $R^{-2}$ fit\\
\hline
isothermal, $\tau_f=6$\,fm & 1.71, 1.67, 1.67, 1.69, 1.71, 1.75 & $0.169\,(m_T+0.066\,{\rm GeV})$\\
isothermal, $\tau_f=8$\,fm & 1.70, 1.67, 1.68, 1.71, 1.74, 1.78 & $0.148\,(m_T+0.021\,{\rm GeV})$\\
isothermal, $\tau_f=10$\,fm & 1.69, 1.67, 1.69, 1.73, 1.75, 1.80 & $0.127\,(m_T-0.003\,{\rm GeV})$\\
isentropic, $\tau_f=8$\,fm & 1.73, 1.84, 1.89, 1.92, 1.94, 1.96 & $0.209\,(m_T-0.128\,{\rm GeV})$\\
isothermal, $q^{-1}=10$\,fm, $\tau_f=8$\,fm & 1.70, 1.68, 1.67, 1.66, 1.67, 1.70 & $0.030\,(m_T+0.247\,{\rm GeV})$\\
\end{tabular}
\end{ruledtabular}
\end{table}

\paragraph{Out and long directions.} The fits above concern the side direction. We now ask whether the source is L\'evy-like in the other two directions as well. They differ from the side direction in what enters the Fourier transform. For $Q$ in the out direction, the mass-shell constraint $Q\cdot K=0$ gives $Q^0=\beta_TQ$ with $\beta_T=K_T/m_T$, so the emission time $t=\tau_f\cosh\eta_s$ contributes. For $Q$ in the long direction at $y=0$, $Q^0=0$, and the coordinate $z=\tau_f\sinh\eta_s$ enters (Appendix~\ref{app:numerics}). Table~\ref{tab:gubserdir} shows the results. The out direction behaves like the side direction, with $\alpha=1.70$--$1.79$. In the long direction the source is close to Gaussian, with $\alpha=1.85$--$1.96$, increasing with $m_T$. This is expected: the power-law tail of the pressure is transverse, while in the longitudinal direction boost invariance leaves only the thermal cutoff $\exp(-m_T\cosh\chi\cosh\eta_s/T_f)$. The source is thus L\'evy-like in the transverse plane only, and an exponent extracted from a one-dimensional or angle-averaged correlation function would lie between the transverse and the longitudinal values. For the weaker flow ($q^{-1}=10$\,fm), the L\'evy scale in the long direction also decreases faster with $m_T$ than the transverse ones, from 8.1 to 3.3\,fm, compared with 8.5 to 5.6\,fm in the side direction.

\paragraph{Origin of the L\'evy-like shape.} Which part of the source makes it L\'evy-like? To answer this we repeat the side-direction fits for the source restricted to $r\le r_{\rm cut}$ (Table~\ref{tab:gubsercut}). The core alone is Gaussian: for $r_{\rm cut}=10$\,fm, $\alpha=2.00$--$2.08$. The exponent then decreases with $r_{\rm cut}$ and reaches its full-source value only at $r_{\rm cut}\approx50$\,fm for the benchmark parameters, and at about 100\,fm for $q^{-1}=10$\,fm. The L\'evy-like shape is therefore produced by the region $r\approx15$--$50$\,fm (benchmark) or $20$--$100$\,fm ($q^{-1}=10$\,fm), i.e.\ by the tail and not by the geometry of the core. Three properties of this region matter for the interpretation.

First, it is not yet the asymptotic slow tail of Subsec.~\ref{sec:gubsertail}. The pressure is already close to its power law there (for the benchmark, $\dd\ln p/\dd\ln r=-5.8$ at $r=20$\,fm and $-5.4$ at 100\,fm, compared with $-16/3$), but the flow is not slow: $v_\perp=0.66$, 0.49 and 0.31 at $r=20$, 30 and 50\,fm. The Boltzmann factor therefore still varies in this region, and the fitted $\alpha\approx1.7$ is a property of this finite range, not an approach to $\alpha_\infty$.

Second, the region is very dilute. For the benchmark, 5--10\% of the pions are emitted from $r>15$\,fm and 1--4\% from $r>20$\,fm, and on $\tau_f$ the density relative to its maximum is $1.2\times10^{-2}$, $1.2\times10^{-3}$ and $7\times10^{-5}$ at $r=20$, 30 and 50\,fm ($6\times10^{-2}$, $9\times10^{-3}$ and $7\times10^{-4}$ for $q^{-1}=10$\,fm). Freeze-out at a common $\tau_f$ and $T_f$ is a strong idealisation for such dilute matter.

Third, the region lies outside the causal envelope of an initial state of nuclear size, $r\lesssim R_A+\tau_f\approx15$\,fm for $R_A\approx7$\,fm (Sec.~\ref{sec:finiterange}). Restricted to this envelope, the source is nearly Gaussian, with $\alpha=1.92$--$1.98$ for the benchmark and $1.99$--$2.05$ for $q^{-1}=10$\,fm.

\begin{table}[t]
\caption{L\'evy fits of the isothermally dressed Gubser flow in the side, out and long directions ($\tau_f=8$\,fm, $T_f=140$\,MeV, pions, full source): $\alpha$ and, in parentheses, $R$ in fm.}
\label{tab:gubserdir}
\begin{ruledtabular}
\begin{tabular}{ccccc}
$q^{-1}$ (fm) & $m_T$ (GeV) & side & out & long\\
\hline
4.3 & 0.2 & 1.70 (5.5) & 1.73 (5.7) & 1.85 (6.6)\\
4.3 & 0.5 & 1.71 (3.6) & 1.77 (3.7) & 1.93 (3.5)\\
4.3 & 0.8 & 1.78 (2.9) & 1.79 (3.1) & 1.96 (2.6)\\
10 & 0.2 & 1.70 (8.5) & 1.73 (8.9) & 1.86 (8.1)\\
10 & 0.5 & 1.66 (6.8) & 1.70 (6.5) & 1.93 (4.4)\\
10 & 0.8 & 1.70 (5.6) & 1.73 (5.5) & 1.96 (3.3)\\
\end{tabular}
\end{ruledtabular}
\end{table}

\begin{table}[t]
\caption{Side-direction L\'evy exponent $\alpha$ of the isothermally dressed Gubser flow ($\tau_f=8$\,fm, $T_f=140$\,MeV, pions) for the source restricted to $r\le r_{\rm cut}$, at $m_T=0.2/0.5/0.8$\,GeV, and the share of the emission that comes from $r>r_{\rm cut}$ at the same $m_T$.}
\label{tab:gubsercut}
\begin{ruledtabular}
\begin{tabular}{ccccc}
 & \multicolumn{2}{c}{$q^{-1}=4.3$\,fm} & \multicolumn{2}{c}{$q^{-1}=10$\,fm}\\
$r_{\rm cut}$ (fm) & $\alpha$ & share (\%) & $\alpha$ & share (\%)\\
\hline
10 & 2.05 / 2.01 / 2.00 & 28 / 29 / 29 & 2.08 / 2.05 / 2.03 & 47 / 52 / 57\\
15 & 1.98 / 1.92 / 1.92 & 9.7 / 6.9 / 5.2 & 2.05 / 2.01 / 1.99 & 24 / 24 / 25\\
20 & 1.90 / 1.84 / 1.85 & 4.1 / 2.2 / 1.3 & 2.01 / 1.95 / 1.93 & 12 / 11 / 10\\
30 & 1.79 / 1.75 / 1.80 & 1.1 / 0.5 / 0.2 & 1.91 / 1.83 / 1.81 & 4.0 / 3.2 / 2.3\\
50 & 1.72 / 1.72 / 1.78 & 0.2 / 0.1 / 0.0 & 1.77 / 1.71 / 1.72 & 0.8 / 0.6 / 0.3\\
100 & 1.70 / 1.71 / 1.78 & 0.0 / 0.0 / 0.0 & 1.71 / 1.67 / 1.70 & 0.1 / 0.1 / 0.0\\
full & 1.70 / 1.71 / 1.78 & -- & 1.70 / 1.66 / 1.70 & --\\
\end{tabular}
\end{ruledtabular}
\end{table}

\subsection{\texorpdfstring{\lPF{}: source functions and L\'evy fits}{lambda-PF: source functions and Levy fits}}\label{sec:obsSolI}
The pressure of \lPF{} has no L\'evy-type tail of its own (Sec.~\ref{sec:solI}), so a power-law tail of its source can only come from the dressing. We examine this in four steps, each treated in one of the paragraphs below:
\begin{enumerate}
\item the isothermal and isentropic dressings, which give no power-law tail and no realistic source size;
\item the fast-tail dressing of Sec.~\ref{sec:hubble}, for which we derive the asymptotic tail exponent and the range of parameters where the tail is L\'evy-stable;
\item a numerical check of this asymptotic exponent;
\item L\'evy fits over the finite range probed by measurements, and the effect of the core acceleration on $\alpha$ and $R$.
\end{enumerate}

\paragraph{Setup.} We freeze out at $\tau=\tau_f$, where $K^\mu\dd\Sigma_\mu=(E-\mathbf k\cdot\mathbf r/t)\,\dd^3r$ as for the Hubble-type solution of Sec.~\ref{sec:hubble}, and fit side-direction L\'evy correlations as in Subsec.~\ref{sec:levyfits}. We use $\tau_f=8$\,fm, $T_f=140$\,MeV and pions throughout. For the equation of state we mostly take $\kappa=5$, for two reasons. First, the Hubble flow, $\lH=0.4$, then lies inside the power ansatz and inside the subsonic window $0.276<\lambda<0.724$, so accelerating ($\lambda<\lH$, $c>1$) and decelerating ($\lambda>\lH$, $c<1$) cores can be compared directly with the Hubble case. Second, unlike for $\kappa=3$, the light-cone exponent $\delta$ of Eq.~\eqref{eq:lightcone} then takes both signs and, as shown in step~2, enters the tail exponent. We take $\lambda=0.3$, 0.4, 0.5 and 0.65, i.e.\ $c=\kappa(1-\lambda)/3=1.17$, 1, 0.83 and 0.58.

\paragraph{Step 1: isothermal and isentropic dressing.} With these choices the source has no power-law tail. In the outer region the flow differs from Hubble flow only by the constant rapidity offset $\zeta$ (Sec.~\ref{sec:solIaccel}), so $K\cdot u\propto r$, while $T$ is either constant (isothermal) or varies only as $(r/\tau)^\delta$ with $|\delta|<1$ (isentropic). By the argument of Sec.~\ref{sec:mismatch}, the Boltzmann factor then cuts off the source: $K\cdot u/T\propto r^{1-\delta}$, so the cutoff is exponential for the isothermal dressing, and $\exp(-c\,r^{1-\delta})$ for the isentropic one, i.e.\ a stretched exponential for $\delta>0$. Moreover, the self-similar flow has no length scale besides $\tau_f$: on $\tau=\tau_f$ the pressure, and for these two dressings also the density, depends on $r$ only through $r/\tau_f$. In the language of Eq.~\eqref{eq:CsL}, there is no geometric size $R_{\rm geom}$, and the source size is set by the flow gradient alone. This is different from the Gubser flow, whose pressure carries the transverse scale $q^{-1}$, which the isothermal dressing passes on to the density. For the four values of $\lambda$ of the setup and pions with $m_T=0.2$--$0.8$\,GeV we find $\alpha\approx1.8$--$1.95$ for both dressings. With the isothermal dressing, $R=9$--$34$\,fm, independent of $m_T$ to within about 12\%. With the isentropic dressing, $R$ depends on $m_T$: from $m_T=0.2$ to $0.8$\,GeV it decreases from 9.2 to 7.6\,fm for $\lambda=0.3$, but it increases from 27 to 32\,fm for $\lambda=0.5$ and from 74 to 120\,fm for $\lambda=0.65$. Since these sources have no power-law tail, we use the full source here, integrated up to $r=3\times10^4$\,fm, where all fits are converged (Appendix~\ref{app:numerics}). Some of them are, however, very broad. For the isentropic dressing with $\lambda=0.65$ and $m_T=0.8$\,GeV, half of $\tilde S(0)$ comes from $r>1.5\times10^3$\,fm, i.e.\ from emission at $t\gtrsim1.5\times10^3$\,fm, almost entirely within a narrow cone around the out direction. There the flow is parallel to the pair momentum, so $K\cdot u$ is smallest, and the outward-growing temperature compensates the Boltzmann factor up to the largest distances. If the source were restricted to $r\le300$\,fm as in step~4, $R$ would decrease with $m_T$ in all cases. For $\lambda\ge0.5$, however, this decrease would be an effect of the truncation, signalled by $\alpha\to2$ (e.g.\ $R=59$ and 37\,fm, with $\alpha=1.93$ and 2.00, for the isentropic $\lambda=0.65$ case). A stronger acceleration of the core gives a steeper velocity gradient and thus a smaller source: with the isothermal dressing, $R\approx9$--$10$\,fm for $c=1.17$, compared to $R\approx23$\,fm for $c=0.83$. These sources are therefore far from the measured ones: $\alpha$ is close to 2, and $R$ is mostly too large and, except for $\lambda=0.3$ with the isentropic dressing, does not decrease with $m_T$ as measured. Two ingredients are missing: a geometric length scale that does not come from the flow gradient, and a power-law tail. Both can be introduced through the dressing.

\paragraph{Step 2: fast-tail dressing and its asymptotic tail.} We use the dressing of the Hubble-type L\'evy solution of Sec.~\ref{sec:hubble}, $\nu=(1+r^2/R_0^2)^{-\beta}$, imposed on $\tau=\tau_f$; this fixes $\nu$ on every streamline. The scale $R_0$, here 5\,fm, plays the role of $R$ in Sec.~\ref{sec:hubble} and of $R_{\rm geom}$ in Eq.~\eqref{eq:CsL}, and $\beta$ sets the power-law tail. Since $n=w^\kappa\nu$ and $T=w/\nu$ with $w\propto(r/\tau)^\delta$, the density then falls as $n\propto r^{\kappa\delta-2\beta}$ and the temperature rises as $T\propto r^{\delta+2\beta}$, so the fast-tail mechanism of Sec.~\ref{sec:hubble} can apply; the condition for it is derived below. \lPF{} thus generalises the Hubble-type L\'evy source to an accelerating or decelerating core, and $\lambda=\lH$ reproduces the Hubble case.

The asymptotic source follows as in Eq.~\eqref{eq:hubbletail}. For $T\gg m$ one has $p_{\rm eq}\propto T^4$, and the fugacity~\eqref{eq:fugp} becomes
\begin{equation}
\frac{p}{p_{\rm eq}(T)}\propto\frac{w^{\kappa+1}}{T^4}=w^{\kappa-3}\nu^4\propto r^{(\kappa-3)\delta-8\beta} .
\end{equation}
In the outer region the flow rapidity is $\eta+\zeta$ with a constant $\zeta$, so $K\cdot u\propto e^{\eta}\propto r/\tau$, and the exponent of the Boltzmann factor behaves as $K\cdot u/T\propto r^{1-\delta-2\beta}$. It approaches a nonzero, angle-dependent constant for $2\beta+\delta=1$ and vanishes for $2\beta+\delta>1$, so a power-law tail is obtained for $2\beta+\delta\ge1$, the analogue of $\beta\ge1/2$ in the Hubble case; for $2\beta+\delta<1$ the tail is a stretched exponential. Under this condition the flux factor is bounded as before, and
\begin{equation}
S\propto r^{-[8\beta-(\kappa-3)\delta]},\qquad \alpha_\infty=8\beta-(\kappa-3)\delta-3 .\label{eq:alphainfI}
\end{equation}
The tail is L\'evy-stable if, in addition, $0<\alpha_\infty<2$, i.e.\ for
\begin{equation}
2\beta+\delta\ge1,\qquad 3<8\beta-(\kappa-3)\delta<5 .\label{eq:betawindowI}
\end{equation}
For $\kappa=3$ the exponent does not depend on $\delta$, and $\alpha_\infty=8\beta-3$ as for Hubble flow; since $\delta\ge0$ for $\kappa=3$, only the lower limit on $\beta$ is relaxed. For $\kappa=5$ and the four values of $\lambda$ of the setup, Table~\ref{tab:solIstable} lists $\delta$, the window~\eqref{eq:betawindowI}, and $\alpha_\infty$ for $\beta=0.55$ and $\beta=0.85$. An accelerating core ($\delta<0$) narrows the window and shifts it to larger $\beta$; a decelerating core widens it. For $\beta=0.55$ all four cases are asymptotically L\'evy-stable, with $\alpha_\infty=0.9$--$1.6$. For $\beta=0.85$ none of them is ($\alpha_\infty=3.3$--$4.0$).

These L\'evy-stable tails have an infinite particle number. To show this, we combine the condition for a finite particle number, $2\beta>2+\kappa\delta$ (Sec.~\ref{sec:solI}), with the condition for a power-law tail, $2\beta\ge1-\delta$, and insert each into $\alpha_\infty-2=8\beta-(\kappa-3)\delta-5$, Eq.~\eqref{eq:alphainfI}. The first condition gives $8\beta>8+4\kappa\delta$, hence
\begin{equation}
\alpha_\infty-2>8+4\kappa\delta-(\kappa-3)\delta-5=3\left[1+(\kappa+1)\delta\right] .
\end{equation}
The second condition gives $8\beta\ge4-4\delta$, hence
\begin{equation}
\alpha_\infty-2\ge4-4\delta-(\kappa-3)\delta-5=-\left[1+(\kappa+1)\delta\right] .
\end{equation}
Whatever the sign of $1+(\kappa+1)\delta$, one of the two right-hand sides is non-negative, so $\alpha_\infty>2$. For every $\kappa$ and $\lambda$, a finite particle number therefore excludes a L\'evy-stable tail; this is the general statement of Sec.~\ref{sec:bound} in explicit form. For Hubble flow ($\delta=0$) it reduces to $\beta>1$ and $\alpha_\infty>5$ (Sec.~\ref{sec:hubble}). Of the eight cases of Table~\ref{tab:solIstable}, only $\lambda=0.3$ with $\beta=0.85$ has a finite particle number ($2\beta=1.7>2+\kappa\delta=1.5$), and its $\alpha_\infty$ is 4.0.

\begin{table}[t]
\caption{Asymptotic tail of \lPF{} with the fast-tail dressing $\nu=(1+r^2/R_0^2)^{-\beta}$ for $\kappa=5$: light-cone exponent $\delta$, Eq.~\eqref{eq:lightcone}; range of $\beta$ with a L\'evy-stable tail, Eq.~\eqref{eq:betawindowI}; and $\alpha_\infty$, Eq.~\eqref{eq:alphainfI}, for $\beta=0.55$ and $\beta=0.85$. In parentheses: $-\dd\ln S/\dd\ln r-3$ from the numerical source in the side direction at $r=10^4$--$3\times10^4$\,fm, for $m_T=0.2$\,GeV. For $\lambda=0.3$ and $\beta=0.55$ one has $2\beta+\delta=1$, and the tail has an angle-dependent prefactor.}
\label{tab:solIstable}
\begin{ruledtabular}
\begin{tabular}{cccccc}
$\lambda$ & $c$ & $\delta$ & L\'evy-stable for & $\alpha_\infty$ ($\beta=0.55$) & $\alpha_\infty$ ($\beta=0.85$)\\
\hline
0.30 & 1.17 & $-0.10$ & $0.55\le\beta<0.60$ & 1.6 (1.60) & 4.0 (3.99)\\
0.40 & 1.00 & 0 & $0.50\le\beta<0.625$ & 1.4 (1.36) & 3.8 (3.80)\\
0.50 & 0.83 & 0.10 & $0.45\le\beta<0.65$ & 1.2 (1.17) & 3.6 (3.60)\\
0.65 & 0.58 & 0.25 & $0.44<\beta<0.69$ & 0.9 (0.89) & 3.3 (3.30)\\
\end{tabular}
\end{ruledtabular}
\end{table}

\paragraph{Step 3: numerical check of the tail.} A direct numerical evaluation of the source confirms Eq.~\eqref{eq:alphainfI}. We evaluate Eq.~\eqref{eq:CF} on $\tau=\tau_f$ for the four values of $\lambda$ and the two values of $\beta$ above, with the full $K_2$-dependent fugacity, for pions with $m_T=0.2$ and $0.8$\,GeV, and determine the local slope $\dd\ln S/\dd\ln r$ in the side direction (Appendix~\ref{app:numerics}). At $r=10^4$--$3\times10^4$\,fm the slopes agree with $-[8\beta-(\kappa-3)\delta]$ to within 0.03 for $\beta=0.85$, and to within 0.16 for $\beta=0.55$ (Table~\ref{tab:solIstable}). The slower convergence for $\beta=0.55$ has a simple origin. For $2\beta+\delta>1$ the exponent of the Boltzmann factor vanishes only as a small power of $r$, $K\cdot u/T\propto r^{-\omega}$ with $\omega=2\beta+\delta-1$. Its logarithmic derivative is $\dd(-K\cdot u/T)/\dd\ln r=\omega\,K\cdot u/T$, so that at finite $r$
\begin{equation}
\frac{\dd\ln S}{\dd\ln r}\simeq-\left[8\beta-(\kappa-3)\delta\right]+\omega\,\frac{K\cdot u}{T} .\label{eq:localslope}
\end{equation}
With $K\cdot u/T$ taken from the solution, Eq.~\eqref{eq:localslope} reproduces the numerical slopes to within about 0.05 already at $r=100$--$300$\,fm, and to within $10^{-3}$ at $r=10^4$--$3\times10^4$\,fm. For small $\omega$ the asymptotic slope is therefore reached only at very large distances, and more slowly for larger $m_T$. For the Hubble case, $\lambda=\lH$ and $\beta=0.85$, the same calculation gives $-6.74$ ($m_T=0.2$\,GeV) and $-6.58$ (0.8\,GeV) at $r=100$--$300$\,fm, in agreement with Sec.~\ref{sec:hubble}.

\paragraph{Step 4: L\'evy fits and the effect of the core acceleration.} The asymptotic tail is reached only at distances far beyond those probed by femtoscopy, so we finally fit the source over the experimentally relevant range, as for the Gubser flow. For $\beta=0.55$ this range has to be specified explicitly. The tail is then so heavy that a sizeable part of the normalisation $\tilde S(0)$ comes from large distances: for $\lambda=0.65$ and $m_T=0.8$\,GeV, 29\% from $r>100$\,fm and 6\% from $r>10^3$\,fm. We therefore restrict the source to $r\le300$\,fm. On $\tau_f=8$\,fm this corresponds to emission at $t\lesssim300$\,fm, and structure at larger $r$ changes the shape of $C(Q)$ only at $Q\lesssim\hbar c/(300\,{\rm fm})\approx0.7$\,MeV, below the resolution of femtoscopic measurements (Sec.~\ref{sec:finiterange}). The cut is not innocuous, however. The emission from $r>300$\,fm also lowers $C(Q)-1$ at all larger $Q$ through the normalisation $\tilde S(0)$, as a halo would, and a fit with the intercept fixed to 1 is sensitive to this. Table~\ref{tab:solIcut} therefore shows how the fits depend on the cut. Only the accelerating core, $\lambda=0.3$, has converged at 300\,fm. For Hubble flow and the decelerating cores $\alpha$ keeps decreasing with the cut, by up to 0.43 between 300\,fm and the full source (for $\lambda=0.65$ and $m_T=0.8$\,GeV). Even the $m_T$ dependence changes: for $\lambda=0.65$, $\alpha$ increases with $m_T$ for $r\le300$\,fm, does not depend on it for $r\le10^3$\,fm, and decreases with it for the full source. A lower limit on $Q$ does not replace the cut either: a fit of the full source with a free intercept over $Q>5$\,MeV gives nearly the full-source values (last column of Table~\ref{tab:solIcut}). For a source with a L\'evy-stable tail the fitted $\alpha$ is thus not defined independently of the range, and the values for $\beta=0.55$ quoted below hold for the 300\,fm cut. The increase of $R$ with $m_T$, in contrast, is found for every cut. For $\beta=0.85$ the fits do not depend on the cut. The fit method is described in Appendix~\ref{app:numerics}. Table~\ref{tab:solI} shows the results for $\kappa=5$. For $\beta=0.55$, whose asymptotic tail is L\'evy-stable, and with the 300\,fm cut, an accelerating core ($c=1.17$) raises $\alpha$ from 1.2 (Hubble) to about 1.4 and nearly removes the increase of $R$ with $m_T$, from $+40\%$ to $+8\%$ between $m_T=0.2$ and $0.8$\,GeV. That the accelerating core gives a larger $\alpha$ than Hubble flow holds for every cut of at least 100\,fm and for the full source (1.34--1.38 compared with 1.13--1.19). A decelerating core has the opposite effect on $R$, whose increase grows to $+78\%$ and $+120\%$ for $c=0.83$ and 0.58. Its effect on $\alpha$ depends on the cut: for the full source $\alpha$ is lower than for Hubble flow (0.75--1.07), but with the 300\,fm cut the difference is smaller and, at $m_T=0.8$\,GeV, not monotonic in $\lambda$. For $\beta=0.85$, where the tail is not L\'evy-stable, the fits still give $\alpha\approx1.7$, almost independently of $\lambda$: as for the Hubble case with the same $\beta$, this is a finite-range property. The acceleration then only moderates the increase of $R$. For $\kappa=3$, where Eq.~\eqref{eq:cwindow} allows stronger acceleration, $\alpha$ for $\beta=0.55$ and $c=1.55$ decreases from 1.10 to 0.90 between $m_T=0.2$ and $0.8$\,GeV (from 1.05 to 0.75 for the full source), opposite to the trend in recent data. In this spherical setup with freeze-out at $\tau=\tau_f$, $R$ increases with $m_T$ for all $\lambda$, unlike in the data. The Hubble-type source of Ref.~\cite{Szentkuti:2026buv}, with an ellipsoidal scaling variable, does not reproduce the hydrodynamic decrease of the radii either (Sec.~\ref{sec:hubble}). Whether other parameter sets or freeze-out hypersurfaces restore it for the fast-tail dressing is an open question (Sec.~\ref{sec:summary}).

\begin{table}[t]
\caption{Side-direction L\'evy fits for \lPF{} with the fast-tail dressing $\nu=(1+r^2/R_0^2)^{-\beta}$ on $\tau_f=8$\,fm ($\kappa=5$, $R_0=5$\,fm, $T_f=140$\,MeV, pions). $\lambda=0.4=\lH$ is Hubble flow; $c=\kappa(1-\lambda)/3$. The source is restricted to $r\le300$\,fm (see text); for $\beta=0.85$ this restriction has no effect, while for $\beta=0.55$ the full source gives $\alpha$ lower by up to 0.43 (Table~\ref{tab:solIcut}). Radii in fm.}
\label{tab:solI}
\begin{ruledtabular}
\begin{tabular}{cccccccc}
 & & \multicolumn{2}{c}{$\beta=0.55$} & & \multicolumn{2}{c}{$\beta=0.85$} & \\
$\lambda$ & $c$ & $\alpha$ ($m_T=0.2/0.8$\,GeV) & $R$ ($0.2/0.8$\,GeV) & & $\alpha$ ($0.2/0.8$\,GeV) & $R$ ($0.2/0.8$\,GeV) & \\
\hline
0.30 & 1.17 & 1.35 / 1.40 & 6.2 / 6.7 & & 1.73 / 1.68 & 3.4 / 4.6 & \\
0.40 & 1.00 & 1.22 / 1.19 & 7.8 / 10.9 & & 1.71 / 1.68 & 3.5 / 5.4 & \\
0.50 & 0.83 & 1.14 / 1.13 & 9.4 / 16.8 & & 1.70 / 1.69 & 3.7 / 6.0 & \\
0.65 & 0.58 & 1.05 / 1.18 & 13.0 / 28.7 & & 1.67 / 1.68 & 3.8 / 6.7 & \\
\end{tabular}
\end{ruledtabular}
\end{table}

\begin{table}[t]
\caption{Dependence of the side-direction L\'evy fits of Table~\ref{tab:solI} for $\beta=0.55$ on the cut $r\le r_{\rm cut}$: $\alpha$ at $m_T=0.2/0.8$\,GeV. ``Full'' is the source without a cut. The last column is a fit of the full source with a free intercept over $Q>5$\,MeV and $C-1>0.1$ (Appendix~\ref{app:numerics}).}
\label{tab:solIcut}
\begin{ruledtabular}
\begin{tabular}{cccccccc}
 & & \multicolumn{4}{c}{$r_{\rm cut}$ (fm)} & & \\
$\lambda$ & $c$ & 50 & 100 & 300 & 1000 & full & free intercept\\
\hline
0.30 & 1.17 & 1.53 / 1.61 & 1.41 / 1.46 & 1.35 / 1.40 & 1.34 / 1.39 & 1.34 / 1.38 & 1.27 / 1.31\\
0.40 & 1.00 & 1.53 / 1.63 & 1.34 / 1.38 & 1.22 / 1.19 & 1.20 / 1.14 & 1.19 / 1.13 & 1.13 / 1.08\\
0.50 & 0.83 & 1.54 / 1.74 & 1.31 / 1.44 & 1.14 / 1.13 & 1.09 / 1.00 & 1.07 / 0.94 & 1.04 / 0.95\\
0.65 & 0.58 & 1.57 / 1.89 & 1.31 / 1.59 & 1.05 / 1.18 & 0.93 / 0.93 & 0.87 / 0.75 & 0.89 / 0.79\\
\end{tabular}
\end{ruledtabular}
\end{table}

\subsection{Comparison of the source tails}\label{sec:tailsummary}
The source tails of the three solutions can now be compared:
\begin{itemize}
\item \emph{Hubble-type L\'evy solution}: power-law tail $S\propto r^{-8\beta}$ (fast tail, fugacity-controlled), L\'evy-stable for $1/2\le\beta<5/8$, Eq.~\eqref{eq:betawindowH}; infinite particle number for $\beta\le1$, Eq.~\eqref{eq:NHubble}.
\item \emph{Dressed Gubser flow}: power-law tail $r^{-16/3}$ (slow tail), L\'evy-like with $\alpha\approx1.7$ in the transverse directions over the experimental range (nearly Gaussian in the long direction), but L\'evy-stable for no dressing at nonzero $K_T$, Eqs.~\eqref{eq:gubserbound} and~\eqref{eq:gubserlowT}.
\item \emph{\lPF{}}: no power-law tail with the isothermal or isentropic dressing; with the fast-tail dressing it hosts the Hubble-type mechanism with an accelerating or decelerating core, and the tail is L\'evy-stable in a range of $\beta$ that depends on $\lambda$, Eq.~\eqref{eq:betawindowI}; a finite particle number implies $\alpha_\infty>2$.
\end{itemize}
Only the dressed Gubser flow combines a L\'evy-like source with radii that decrease with $m_T$ as in hydrodynamics; the fast-tail sources are L\'evy-stable, but only with an infinite particle number, and their radii do not show this decrease.

%=====================================================================
\section{Towards non-sphericality and elliptic flow}
\label{sec:nonsph}
%=====================================================================

\subsection{Hubble-type solutions}
Ellipsoidal scaling variables, as in Eq.~\eqref{eq:hubblelevy} with $v_x\neq v_y$, generate elliptic flow through the anisotropic temperature and density profiles even though the flow itself is isotropic~\cite{Csanad:2009wc}. Multipole generalisations give higher harmonics~\cite{Csanad:2014dpa}. The L\'evy member therefore allows a simultaneous calculation of $v_2$ and L\'evy HBT~\cite{Szentkuti:2026buv}.

\subsection{Dressing and anisotropy}
An anisotropic dressing $\nu(\theta,\varphi)$ of an isotropic flow is allowed by the theorem of Sec.~\ref{sec:general}. With the isothermal dressing on $\tau_f$, however, $n=p/T_f$ inherits the isotropy of $p$, so no anisotropy arises. Relaxing isothermality produces a ``kinematic'' $v_2$ from anisotropic $n$ and $T$ combined with radial flow, as in the Hubble case. A \emph{dynamical} $v_2$, generated by pressure gradients, requires an anisotropic flow solution.

\subsection{Eccentric Gubser flow}
Gubser and Yarom constructed linearised perturbations of Gubser flow in de Sitter space, including eccentric (azimuthally dependent) modes~\cite{Gubser:2010ui}. Since the dressing theorem holds for any flow, these perturbed solutions can be dressed in the same way. This allows $v_2$, azimuthally sensitive HBT radii, and the azimuthal dependence of $\alpha$ to be computed consistently; we leave this to future work.

\subsection{Ellipsoidal extension of \lPF{}}
The spherical ansatz generalises to $\Phi=t^\lambda\big[F(s)+\epsilon\,F^{(2)}(s)P_2(\cos\vartheta)+\dots\big]$, where $F$ is the spherical solution and $\epsilon$ a small amplitude. At first order in $\epsilon$, Eq.~\eqref{eq:PhiPDE} becomes a linear ODE for $F^{(2)}$. Regularity requires $F^{(2)}\propto s$ at the centre, and the light-cone or sonic conditions give an eigenvalue-type relation. The light-cone exponent of $F^{(2)}$ determines whether the eccentricity grows or decays. Deriving and solving this equation is left for future work, as are fully ellipsoidal self-similar potentials, $\Phi=t^\lambda F(s_x,s_y,s_z)$, which lead to a PDE in three similarity variables.

\subsection{Beyond potential flow}
Rotating and vortical flows require a Clebsch-type representation instead of Eq.~\eqref{eq:potential}. The dressing theorem itself does not rely on irrotationality.

%=====================================================================
\section{Summary and outlook}
\label{sec:summary}
%=====================================================================

We addressed the L\'evy hydro puzzle with exact solutions of relativistic hydrodynamics. Our main results are the following.

\begin{enumerate}
\item For $\varepsilon=\kappa p$ and $p=nT$, flow and pressure form a closed system. Density and temperature are a free comoving dressing of any flow solution, which generalises the freedom known from Hubble-type solutions.
\item On the freeze-out hypersurface the source is bounded by the pressure, Eq.~\eqref{eq:bound}, so the source tail can never be heavier than the pressure tail. If the temperature tends to zero in the tail, the bound still holds at nonzero $K_T$, with a constant that diverges as $K_T\to0$.
\item There are two simple routes to power-law source tails at fixed pair momentum. In the fast-tail route (Hubble type), the temperature grows at least linearly and the fugacity in the Cooper--Frye formula suppresses the emission from the hot tail~\cite{Szentkuti:2026buv}; the Hubble-type source is L\'evy-stable in the strict, asymptotic sense for $1/2\le\beta<5/8$. Its particle number, however, is finite only for $\beta>1$, and in general a L\'evy-stable tail in the fast-tail route requires an infinite particle number (Sec.~\ref{sec:bound}). In the slow-tail route (accelerating), the temperature is uniform and the proper acceleration falls off as $1/r$. For a quasi-static tail the coefficient is $(3+\alpha)/(\kappa+1)$; an expanding tail adds a contribution of the same order, e.g.\ $a\to2/r$ for the dressed Gubser flow. Companion solutions (Appendix~\ref{app:companion}), a self-similar tail family and an exact Rindler atmosphere~\cite{Borshch:2007uf,Nagy:2007xn}, realise the $1/r$ acceleration explicitly.
\item The family of $\lambda$ potential flows (\lPF) is relativistic, accelerating and regular, with Hubble flow as a special case. Its acceleration is harmonic in the core, $a\simeq c(c-1)\,r/t^2$ with $c=\kappa(1-\lambda)/3$ bounded by the subsonic window, and tends to a constant on $\tau=$\,const in the outer region instead of falling off as $1/r$. It therefore produces no L\'evy tail of its own, but with a fast-tail dressing it generalises the Hubble-type L\'evy source to accelerating cores; the tail is then L\'evy-stable for a range of $\beta$ that depends on $\lambda$, Eq.~\eqref{eq:betawindowI}, as confirmed numerically. A finite particle number, however, implies $\alpha_\infty>2$ for every $\kappa$ and $\lambda$, and for the L\'evy-stable tails the fitted $\alpha$ depends on the range of the source that is included in the fit. In the spherical setup studied here, however, the radii increase with $m_T$.
\item The isothermally dressed Gubser flow is an exact, accelerating solution with transverse and longitudinal expansion and a power-law source tail; being boost invariant and azimuthally symmetric, it depends on two variables only. In the transverse (side and out) directions it gives L\'evy-like fits with $\alpha\approx1.7$, and $R^{-2}$ in the side direction is linear in $m_T$; in the long direction the source is nearly Gaussian ($\alpha=1.85$--$1.96$). The L\'evy-like shape is produced by the dilute outer region, $r\approx15$--$50$\,fm for the benchmark parameters, which emits less than 10\% of the pions. For the benchmark parameters $\alpha$ increases with $m_T$, but the transverse flow is too strong for realistic spectra. For a weaker flow ($q^{-1}=10$\,fm) the spectra are nearly realistic, $\alpha=1.66$--$1.70$ with a weaker increase at higher $m_T$, and $R^{-2}$ stays linear. Kaons and pions with the same $m_T$ give similar sources only for moderate flow or at large $m_T$. The tail exponent is fixed by the Gubser solution at $\alpha_\infty=10/3$, so these fits describe a finite range, and no dressing makes the tail L\'evy-stable at nonzero $K_T$. The isothermal dressing implies a hot, dilute periphery at early times, which does not affect the perfect-fluid dynamics.
\end{enumerate}

Table~\ref{tab:requirements} compares the solutions with the four requirements of Sec.~\ref{sec:intro}. None of them meets all four. The new solutions are accelerating but depend on two variables only: on $t$ and $r$ for \lPF{}, on $\tau$ and $r$ for the dressed Gubser flow. The Hubble-type L\'evy solution is genuinely 1+3 dimensional but accelerationless. With suitable dressings, all three give L\'evy-like or L\'evy-stable sources, but none reproduces all measured trends.

\begin{table}[t]
\caption{The requirements of Sec.~\ref{sec:intro}, (i) relativistic, (ii) accelerating, (iii) genuinely 1+3 dimensional, and (iv) realistic observables including L\'evy-like sources, for the solutions discussed in this paper.}
\label{tab:requirements}
\begin{ruledtabular}
\begin{tabular}{lccll}
Solution & (i) & (ii) & (iii) & (iv)\\
\hline
\parbox[t]{0.2\linewidth}{\raggedright Hubble-type L\'evy, Eq.~\eqref{eq:hubblelevy}\strut} & yes & no & yes & \parbox[t]{0.42\linewidth}{\raggedright partly: L\'evy-stable for $1/2\le\beta<5/8$ (infinite particle number for $\beta\le1$), L\'evy-like fits; pion $T_{\rm eff}\approx0.3$--$0.5$\,GeV; radii without hydrodynamic $m_T$ scaling~\cite{Szentkuti:2026buv}\strut}\\
\parbox[t]{0.2\linewidth}{\raggedright \lPF{}, fast-tail dressing\strut} & yes & yes & \parbox[t]{0.16\linewidth}{\raggedright no (spherical)\strut} & \parbox[t]{0.42\linewidth}{\raggedright partly: L\'evy-stable for a $\lambda$-dependent range of $\beta$ (with an infinite particle number); $R$ increases with $m_T$\strut}\\
\parbox[t]{0.2\linewidth}{\raggedright \lPF{}, isothermal or isentropic dressing\strut} & yes & yes & \parbox[t]{0.16\linewidth}{\raggedright no (spherical)\strut} & \parbox[t]{0.42\linewidth}{\raggedright no: no power-law tail; $R$ nearly independent of $m_T$ (isothermal) or $m_T$-dependent (isentropic)\strut}\\
\parbox[t]{0.2\linewidth}{\raggedright Gubser, isothermal dressing\strut} & yes & yes & \parbox[t]{0.16\linewidth}{\raggedright no (boost invariant, azimuthally symmetric)\strut} & \parbox[t]{0.42\linewidth}{\raggedright partly: $\alpha\approx1.7$ in the transverse directions and linear $R^{-2}(m_T)$; nearly Gaussian in the long direction; nearly realistic spectra for $q^{-1}\approx10$--$12$\,fm; $\alpha_\infty=10/3$; species $m_T$ scaling only approximate\strut}\\
\end{tabular}
\end{ruledtabular}
\end{table}

\paragraph{Answer to the L\'evy hydro puzzle.} Within perfect-fluid hydrodynamics with Cooper--Frye freeze-out, a L\'evy-like source and the hydrodynamic $m_T$ scaling of the radii are compatible. The isothermally dressed Gubser flow gives both in the transverse directions, with $\alpha\approx1.7$ and $R^{-2}$ linear in $m_T$, also for a transverse flow that gives nearly realistic spectra. Hydrodynamics does not, however, create the power-law tail. In the slow-tail route the source tail is the pressure tail, which has to be present already in the initial state and is carried along by the $1/r$ acceleration; an initial state of finite extent cannot develop it, since the fluid stays inside its causal envelope (Sec.~\ref{sec:finiterange}). For the dressed Gubser flow this can be quantified: the L\'evy-like shape is produced by the dilute region $r\approx15$--$50$\,fm, and the source restricted to the causal envelope of an initial state of nuclear size, $r\lesssim15$\,fm, is nearly Gaussian, with $\alpha=1.92$--$1.98$ (Sec.~\ref{sec:levyfits}). In the fast-tail route the tail is put in through the dressing, i.e.\ through the initial entropy per particle. Within the solutions studied here, the slow-tail route reproduces the observed trends, $\alpha<2$ increasing with $m_T$ (for nearly realistic flow only above $m_T\approx0.5$\,GeV) and linear $R^{-2}(m_T)$, but with an asymptotic index outside the stable range, whereas the fast-tail route gives L\'evy-stable tails only together with an infinite particle number, radii that do not decrease with $m_T$, and a power-law spectrum at high $m_T$. None of the solutions studied here thus combines a finite particle number with a L\'evy-stable tail. The slow-tail route allows this combination in principle, since a pressure tail $\propto r^{-(3+\alpha)}$ is integrable, but the tail index of the Gubser flow is fixed at $10/3$. Experiments probe a finite range, where the slow-tail route suffices. The coexistence of L\'evy-like sources and hydrodynamic scaling is therefore not a contradiction, but it shifts the question to the origin of the power-law tail of the initial state, which hydrodynamics alone does not explain.

Subsequent research could then include a quantitative comparison with data (spectra, $\alpha$, $R$, $\lambda$) for pions and kaons. The freeze-out on other hypersurfaces could also be investigated, e.g.\ at constant temperature for dressings other than the isothermal one (for which $\tau=\tau_f$ is itself an isotherm), and the inclusion of resonance decays. It would be interesting to see whether non-spherical parameter sets or other freeze-out hypersurfaces restore the hydrodynamic decrease of the radii with $m_T$ for the fast-tail dressing (Sec.~\ref{sec:obsSolI}). The continuation of the transonic solutions of \lPF{} through the nodal critical point, above and below the subsonic window (Sec.~\ref{sec:solI}) is also worth following up. Furthermore, the dynamical generation of $v_2$ through eccentric perturbations of the Gubser flow could be investigated, including an estimate of $v_2/\varepsilon_2$, and the ellipsoidal extension of \lPF{} (Sec.~\ref{sec:nonsph}). Finally, one could construct a regular core joined to a $1/r$-accelerating tail beyond the Gubser flow, so that the tail exponent becomes a free parameter. Within the routes discussed here, this is the way to a L\'evy-stable tail with a finite particle number. These topics fall out of the scope of present manuscript.

\begin{acknowledgments}
The authors thank T. Csörgő, G. Kasza, B. Tomá{\v{s}}ik for the insightful discussions. This work was supported by the NKFIH grants NKKP ADVANCED 152097 and 2026-4.1.1-MISSZIÓK-2026-00005.

\end{acknowledgments}

\section*{Use of AI tools}
The AI assistant Claude Opus 5.5 was used to check analytic derivations, develop parts of the numerical code, and edit the manuscript text. Subsequently, Claude Fable 5.2 and ChatGPT version 6 were used for additional cross-checking selected analytical derivations and numerical results, and for reviewing the logical consistency of the manuscript. The authors directed and critically reviewed these checks, independently verified the relevant results, and take full responsibility for the scientific content and conclusions of this work.

\appendix

%=====================================================================
\section{Conventions and Milne-coordinate identities}
\label{app:conv}
%=====================================================================
In the Milne coordinates $(\tau,r,\varphi,\eta_s)$ the non-vanishing Christoffel symbols are
\begin{equation}
\Gamma^\tau_{\eta\eta}=\tau,\quad \Gamma^\eta_{\tau\eta}=\frac1\tau,\quad \Gamma^r_{\varphi\varphi}=-r,\quad \Gamma^\varphi_{r\varphi}=\frac1r ,
\end{equation}
and $\sqrt{-g}=\tau r$. For a boost-invariant, azimuthally symmetric flow the two independent components of $\nabla_\mu T^{\mu\nu}=0$ are
\begin{align}
\frac{1}{\tau r}\left[\partial_\tau(\tau rT^{\tau\tau})+\partial_r(\tau rT^{\tau r})\right]+\frac p\tau&=0,\\
\frac{1}{\tau r}\left[\partial_\tau(\tau rT^{\tau r})+\partial_r(\tau rT^{rr})\right]-\frac pr&=0 .
\end{align}

%=====================================================================
\section{Companion solutions with $1/r$ acceleration}
\label{app:companion}
%=====================================================================

\paragraph{Self-similar flows with power-law initial data.} Extending the similarity variable $\xi=r/t$ to $0\le\xi<\infty$, with
\begin{equation}
n=t^{-a_n}N(\xi),\qquad T=t^{-a_T}\Theta(\xi),\qquad v=V(\xi),
\end{equation}
describes initial data at rest with $n_0\propto r^{-a_n}$ and $T_0\propto r^{-a_T}$. The velocity obeys a single ODE depending only on the pressure exponent $a_p=a_n+a_T$, as the dressing theorem requires:
\begin{equation}
V'=\frac{\dfrac{2(\kappa+1)(1-V\xi)V}{\kappa\xi}-a_p(1-V^2)}{\gamma^2(\kappa+1)\left[(V-\xi)^2-(1-V\xi)^2/\kappa\right]} .
\end{equation}
The functions $N$ and $\Theta$ follow by quadrature. The regular centre has $V\simeq c\,\xi$ with $c=\kappa a_p/[3(\kappa+1)]$, and the tail branch has
\begin{equation}
v\simeq\frac{a_p\,t}{(\kappa+1)\,r}\qquad(r\gg t),
\end{equation}
which is precisely the $1/r$ acceleration of Eq.~\eqref{eq:criterion}. The regular centre branch is the regular branch of \lPF{} (Sec.~\ref{sec:solI}) with $\lambda=1-a_p/(\kappa+1)$. First, the scaling agrees: $p\propto t^{-a_p}$ here, and $p=w^{\kappa+1}\propto t^{(\lambda-1)(\kappa+1)}$ for \lPF{}. Second, the central slopes agree, $\kappa a_p/[3(\kappa+1)]=\kappa(1-\lambda)/3$. Third, every radial flow that obeys Euler's equation is a potential flow. Equation~\eqref{eq:euler} gives $wDu_\mu=\Delta_{\mu\nu}\partial^\nu w$, hence $D(wu_\mu)=\partial_\mu w$, so the vorticity $\omega_{\mu\nu}=\partial_\mu(wu_\nu)-\partial_\nu(wu_\mu)$ satisfies $u^\nu\omega_{\nu\mu}=D(wu_\mu)-u^\nu\partial_\mu(wu_\nu)=\partial_\mu w-\partial_\mu w=0$, where $u^\nu\partial_\mu u_\nu=0$ was used. In a radial flow $\omega_{tr}$ is the only component, so $u^t\omega_{tr}=0$ implies $\omega_{tr}=0$, i.e.\ $wu_\mu=\partial_\mu\Phi$. Consequently, inside the window of Sec.~\ref{sec:solI} the centre branch reaches the light cone subsonically, with $v\to1$. For $\kappa=3$ this holds for $1.69<a_p<6.31$, which includes the whole L\'evy range $3<a_p<5$; $a_p=4$ gives exactly Hubble flow, $V=\xi$. The tail branch is exactly $V=1/\xi$, the Rindler atmosphere of Eq.~\eqref{eq:rindler}, for $a_p=\kappa+1$. For $a_p>\kappa+1$ it reaches $V\to1$ as $\xi\to1^+$, and for $a_p<\kappa+1$ it crosses the light cone and folds inside it, just inside for $a_p$ close to $\kappa+1$ (e.g.\ at $\xi\approx0.998$ for $\kappa=3$, $a_p=3.5$, but at $\xi\approx0.96$ for $a_p=2.5$). The two branches therefore cannot be joined smoothly: inside the window the centre branch ends on the light cone with $v\to1$, and outside the window it reaches the sonic line (cases (ii) and (iii) of Sec.~\ref{sec:solI}). Pure power-law data also carry infinite energy near $r=0$. The self-similar solution is therefore the intermediate asymptotics of the tail, not a global solution.

\paragraph{Exact spherical Rindler atmosphere.} For $r>|t|$, with $\rho=\sqrt{r^2-t^2}$, the following is an exact solution for any $\kappa$~\cite{Borshch:2007uf} (see also Ref.~\cite{Nagy:2007xn}):
\begin{equation}
u^\mu=\frac{(r,\,t)}{\rho},\qquad
p=p_0\left(\frac{\rho_0}{\rho}\right)^{\kappa+1}\left(\frac\rho r\right)^{2(\kappa+1)/\kappa},\qquad
\nu=\nu(\rho).\label{eq:rindler}
\end{equation}
The proper acceleration is exactly $1/\rho$. Choosing $\nu\propto1/\rho$ gives $T=T_0(\rho/r)^{2/\kappa}$, uniform at $t=0$, with $n\propto r^{-(\kappa+1)}$ there. This corresponds to a L\'evy index $\alpha=\kappa-2$; $\kappa=3$ gives a Cauchy-like tail ($\alpha=1$). The solution is singular on the light cone $r=|t|$ and can serve only as a tail. It is the $\lambda=0$ member of the self-similar potential flows of Sec.~\ref{sec:solI}, $\Phi=F(r^2/t^2)$ with $F'$ given by Eq.~\eqref{eq:Flam0}, continued to $r>t$ (Sec.~\ref{sec:central}).

%=====================================================================
\section{Second-order coefficient of \lPF{}}
\label{app:F2}
%=====================================================================
\begin{equation}
F_2=\frac{\kappa^2\lambda(\lambda-1)}{1080}\Big[2\kappa^2\lambda^2-4\kappa^2\lambda+2\kappa^2-8\kappa\lambda^2+28\kappa\lambda-20\kappa+15\lambda^2-69\lambda+72\Big]F_0 .
\end{equation}

%=====================================================================
\section{Numerical methods}
\label{app:numerics}
%=====================================================================
\paragraph{L\'evy fits.} For the Gubser flow, the transverse source is evaluated on a grid of half-width 160\,fm and spacing 0.08\,fm and projected onto the side axis. $y(Q)=C(Q)-1$ follows from its cosine transform at 600 logarithmically spaced values of $Q$ between 0.5\,MeV and 0.5\,GeV. $\alpha$ and $R$ are determined by an unweighted linear regression of $\ln(-\ln y)$ on $\ln Q$ over $0.1<y<0.9$, i.e.\ with the intercept fixed to $\lambda=1$. The fit window $0.2<y<0.8$ ($0.05<y<0.95$) changes $\alpha$ by $-0.02$ ($+0.02$). A fit with a free intercept gives $\alpha$ lower by 0.08--0.16, with an intercept of 1.02--1.05. For all these choices $\alpha$ increases with $m_T$ above $m_T\approx0.5$\,GeV.

\paragraph{Directions and cuts for the dressed Gubser flow.} For Tables~\ref{tab:gubserdir} and~\ref{tab:gubsercut} the Fourier transform of the source is reduced to a single radial integral, because both angular integrals can be done analytically. We write $A=m_T\cosh\chi/T_f$ and $b=K_T\sinh\chi/T_f$, so that before the $\eta_s$ integration the source~\eqref{eq:Sperp} is $\propto n\cosh\eta_s\exp(-A\cosh\eta_s+b\cos\varphi)$, with $\varphi$ measured from $\mathbf K_T$. Three integrals are needed. The first is $\int\dd\eta_s\cosh\eta_s\,e^{-z\cosh\eta_s}=2K_1(z)$ for ${\rm Re}\,z>0$. The second follows from $\int\dd\eta_s\,e^{-A\cosh\eta_s+ic\sinh\eta_s}=2K_0(\sqrt{A^2+c^2})$ by differentiation with respect to $A$: $\int\dd\eta_s\cosh\eta_s\,e^{-A\cosh\eta_s}\cos(c\sinh\eta_s)=2A\,K_1(\sqrt{A^2+c^2})/\sqrt{A^2+c^2}$. The third is the azimuthal integral $\int\dd\varphi\,e^{a\cos\varphi+a'\sin\varphi}=2\pi I_0(\sqrt{a^2+a'^2})$, valid for complex $a$ and $a'$. The phase of the Fourier transform is $Qr\sin\varphi$ for $Q$ in the side direction, $Q(\beta_Tt-r\cos\varphi)$ with $t=\tau_f\cosh\eta_s$ in the out direction, and $Qz$ with $z=\tau_f\sinh\eta_s$ in the long direction. Hence, up to common factors,
\begin{align}
\tilde S_{\rm side}(Q)&\propto\int r\,\dd r\,n\,K_1(A)\,I_0\!\left(\sqrt{b^2-Q^2r^2}\right),\\
\tilde S_{\rm out}(Q)&\propto\int r\,\dd r\,n\,K_1(A-iQ\beta_T\tau_f)\,I_0(b-iQr),\\
\tilde S_{\rm long}(Q)&\propto\int r\,\dd r\,n\,\frac{A\,K_1\!\left(\sqrt{A^2+Q^2\tau_f^2}\right)}{\sqrt{A^2+Q^2\tau_f^2}}\,I_0(b),
\end{align}
where $I_0$ of an imaginary argument is $J_0$. The radial integrals are evaluated with the midpoint rule (spacing 0.01\,fm, $r\le3000$\,fm) and, for Table~\ref{tab:gubsercut}, restricted to $r\le r_{\rm cut}$. The fit is the same as above. For the full source, the side-direction results agree with those of the two-dimensional grid to within 0.001 in $\alpha$.

\paragraph{Spectra and kaon sources of the dressed Gubser flow.} The spectrum~\eqref{eq:spectrum} is integrated over $r$ by adaptive quadrature up to $r=400$\,fm, with exponentially scaled Bessel functions. Effective temperatures are $-1/$slope of linear fits of $\ln E\,\dd N/\dd^3K$ versus $m_T$ over $m_T-m=0.1$--$1.0$\,GeV. For the L\'evy fits of the $q$ scan the transverse grid is scaled with $q^{-1}$: half-width $40\,q^{-1}$ and spacing $0.02\,q^{-1}$. Kaons are treated exactly like pions, with $m=0.494$\,GeV in Eq.~\eqref{eq:Sperp}.

\paragraph{Source tails of \lPF.} We use the regular branch of Sec.~\ref{sec:solI}, integrated up to $r/\tau\approx10^4$, and take $w$ and the flow rapidity on $\tau_f=8$\,fm from it. The dressing is $\nu=(1+r^2/R_0^2)^{-\beta}$ with $R_0=5$\,fm, and $T=T_f\,w/[w(0)\,\nu]$ with $T_f=140$\,MeV. The source~\eqref{eq:CF} is evaluated per unit volume with the full $K_2$-dependent fugacity, for pions at $y=0$ with $\mathbf K_T$ along the out direction. Local slopes are linear fits of $\ln S$ versus $\ln r$ in the side direction over $r=10^2$--$3\times10^2$, $10^3$--$3\times10^3$ and $10^4$--$3\times10^4$\,fm; the angle-averaged source approaches the same slopes.

\paragraph{L\'evy fits of \lPF.} For the fits of Table~\ref{tab:solI}, $\tilde S(Q)$ for $Q$ in the side direction is obtained from a Legendre expansion of $S$ in the cosine $\mu$ of the angle to $\mathbf K_T$. Writing $e^{i\mathbf Q\cdot\mathbf r}$ as a sum over $i^\ell(2\ell+1)j_\ell(Qr)P_\ell$ and averaging over the azimuth, the side direction ($\mathbf Q\perp\mathbf K_T$) picks up the factor $P_\ell(0)$, which vanishes for odd $\ell$ and equals $(-1)^{\ell/2}(\ell-1)!!/\ell!!$ for even $\ell$. Since $i^\ell(-1)^{\ell/2}=1$ for even $\ell$,
\begin{equation}
\tilde S(Q)=2\pi\sum_{\ell\,{\rm even}}(2\ell+1)\frac{(\ell-1)!!}{\ell!!}\int_0^{r_{\max}}r^2\,\dd r\,S_\ell(r)\,j_\ell(Qr),\qquad S_\ell(r)=\int_{-1}^1\dd\mu\,S(r,\mu)P_\ell(\mu),
\end{equation}
with $\ell\le60$ and $r_{\max}=300$\,fm. The fit is the same as for the Gubser flow. For Table~\ref{tab:solIcut} the radial integrals are cut at $r_{\rm cut}$ instead; the full source is integrated up to $3\times10^4$\,fm, and $\tilde S(0)$ is extended analytically beyond that with the local power law of the source. The fit with a free intercept is a least-squares fit of $\Lambda\,e^{-|QR|^\alpha}$ to $C(Q)-1$ over $Q>5$\,MeV and $C-1>0.1$; it gives $\Lambda=0.92$--$1.02$. For the isothermal and isentropic dressings of step~1 the same expansion is applied to the full source, with $r_{\max}=3\times10^4$\,fm. Reducing $r_{\max}$ to $10^4$\,fm changes none of these fits by more than 0.002 in $\alpha$ and 0.1\,fm in $R$, and $3\times10^3$\,fm changes only the isentropic case with $\lambda=0.65$ at $m_T=0.8$\,GeV ($\alpha=1.88$ and $R=106$\,fm instead of 1.80 and 120\,fm).

%=====================================================================
\section{Code availability}
\label{app:code}
%=====================================================================
The symbolic and numerical checks, the numerical integration of \lPF{}, and the source and fitting codes, including the scripts that produce all figures and tables of this paper, are available at \href{https://github.com/csanadm/LevyHydroSolutionCodes/}{github.com/csanadm/LevyHydroSolutionCodes}.

\bibliography{references}

\end{document}